\documentclass[twocolumn,showpacs,preprintnumbers,amsmath,amssymb,pre]{revtex4-1}

\usepackage{color}    
\usepackage{graphicx}
\usepackage{dcolumn}
\usepackage{bm}
\usepackage{subfigure}
\usepackage{amssymb}
\usepackage{multirow}
\usepackage{amsmath}
\graphicspath{{plots/}}

\renewcommand{\vec}[1]{\mathbf{#1}}

\begin{document}

\title{Structural characteristics of strongly coupled ions in a dense quantum plasma}

\author{ Zh.~A.  Moldabekov$^{1, 2}$, S.~ Groth$^{1}$, T.~Dornheim$^{1}$, H.~K\"ahlert$^{1}$, M. Bonitz$^{1}$, and T. S.  Ramazanov$^{2}$}

\affiliation{
 $^1$Institut f\"ur Theoretische Physik und Astrophysik, Christian-Albrechts-Universit\"at zu Kiel,
 Leibnizstra{\ss}e 15, 24098 Kiel, Germany}
 \affiliation{
 $^2$Institute for Experimental and Theoretical Physics, Al-Farabi Kazakh National University, 71 Al-Farabi str.,  
  050040 Almaty, Kazakhstan
  }

\begin{abstract}
The structural properties of strongly coupled ions in dense plasmas with moderately to strongly degenerate electrons are investigated in the framework of the 
one-component plasma model of ions interacting through a screened pair interaction potential. 
Special focus is put on the description of the electronic screening in the Singwi-Tosi-Land-Sj\"olander (STLS) approximation.  Different cross-checks 
and analyses using ion potentials obtained from ground-state quantum Monte Carlo data, the random phase approximation (RPA), 
and existing analytical models are presented for the computation of the structural properties, such as the pair distribution and the static structure factor, of strongly coupled ions. 
The results are highly sensitive to the features of the screened pair interaction potential.
This effect is particularly visible in the static structure  factor.  
The applicability range of the screened potential computed from STLS is identified in terms of density and temperature of the electrons. 
It is demonstrated that at $r_s>1$, where $r_s$ is the ratio of the mean inter-electronic distance to the Bohr radius, electronic correlations beyond RPA have a non-negligible effect on the structural properties.
Additionally, the  applicability of the hypernetted chain approximation for the calculation of the structural properties  using the screened pair interaction potential is analyzed employing the effective coupling parameter 
 approach.

\end{abstract}

\pacs{xxx}

\maketitle
\section{Introduction}

  Dense plasmas with different temperatures of ions and electrons are realized in experiments on inertial confinement fusion (ICF) \cite{Zastrau, Hurricane, Cuneo, Gomez}. 
 During compression of the target by a laser or intense charged particle beams \cite{Hoffmann1, Boris, Kawata}  electrons get heated first, followed by the subsequent thermalization with an ionic subsystem.  
 In such plasmas, full equilibration can be finally reached due to the energy  exchange between electrons and ions. However, due to the large ion to electron mass ratio, the temperature equilibration is rather slow. 
Depending on the plasma density and initial values of the temperatures of  electrons and ions, the thermalization time is in the range from $\sim10^3~{\rm fs}$ to $\sim 10^5~{\rm fs}$ \cite{Hartley, White2014, MRE2017, Gericke, Glosli}. This time is much larger than the characteristic 
time scale of the ionic subsystem of dense plasmas, which is $\sim \omega_{\rm pi}^{-1}$, where $\omega_{\rm pi}$ denotes the ion plasma frequency, and this time increases with the plasma coupling strength \cite{ohde_pop_96}.
This results in a transient stationary nonequilibrium state of a dense plasma with relatively cold strongly coupled ions and hot ideal or weakly coupled electrons \cite{Ravasio, Clerouin, Lyon, Ma2014}.
Another reason for the interest in a dense quantum plasma with strongly coupled ions is  the study of the properties of extreme states of matter upon laser compression of materials \cite{ Fortov_book, Ravasio, TahirCPP, Garcia} and laboratory astrophysics \cite{Tahir2011, Tahir2017}. 

The strong coupling within the ionic subsystem can be detected by analyzing the static structure factor $S(k)$, which is measured using the X-ray Thomson scattering  technique \cite{book_david}.
 For instance, in a recent experiment on the laser shock-compressed aluminum, the product of the Fourier transform of the electron density and the static structure factor of strongly coupled ions
 was measured via X-ray Thomson scattering \cite{Ma2014}. 
 
 Motivated by the  experimental realization of  dense two-temperature plasmas \cite{Ma, PhysRevLett.98.065002, PhysRevLett.102.115001}, in this paper we consider  a fully ionized quantum plasma consisting of
 weakly non-ideal partially degenerate electrons and strongly coupled classical ions.
  The theoretical description of such plasmas is challenging due to the simultaneous effect of non-ideality, electron quantum degeneracy, thermal excitation, and mostly because of the out-of-equilibrium condition. 
    At present, there exists disagreement on peculiarities of the ionic structure factor of dense plasmas. For instance,
    in order to find an explanation of the observed structural characteristics of the ions, Fletcher \textit{et al.}~\cite{Fletcher} and Ma \textit{et al.}~\cite{Ma} proposed an effective ion-ion interaction potential consisting of
a Yukawa potential with an additional short-range repulsive potential. This model was questioned by Cl\'erouin \textit{et al.}~\cite{Clerouin}, who investigated the structural characteristics of two-temperature dense plasmas by a molecular dynamics simulation of ions based on 
 a Thomas-Fermi density functional theory treatment of the electrons. Later, Harbour \textit{et al.} \cite{Harbour} investigated  
 the compressibility, phonons, and electrical conductivity of warm dense matter on the basis of an improved neutral-pseudoatom model and also found disagreement with the conclusions of Refs.~\cite{Fletcher, Ma}.

Due to the high complexity of the study of dense plasmas out-of-equilibrium, a careful analysis of both the experimental and the simulation data by performing comparison to well defined models with clear approximations is indispensable. 
 Such a comparative analysis helps to identify the inner machinery of the microscopic processes, which cannot be observed directly in measurements.  
 In fact, previous works on dense plasmas and warm dense matter hugely benefited from such comparisons. 
 For instance, the analysis of the dynamical conductivity using the
 Mermin dielectric function helped to identify the non-Drude-like behavior induced by electron excitations in the
conduction band \cite{WittePRL}. The results from the one-component  Coulomb  plasma model (OCP) and the Yukawa one-component plasma model (YOCP) 
were used for the  analysis of the structural properties of warm dense matter \cite{Clerouin, Wunsch}, where  
 the comparison with the OCP results helped to understand the role of screening, whereas comparison with the YOCP results is needed for having a picture about higher order electronic correlation effects. 
 
 Regarding structural properties, clearly, it will be very useful to compare data from \textit{ab-initio} simulations to  results obtained on the basis of
  improved ion potentials in the framework of the linear response theory \cite{CPP2017}.  
Indeed, if the quantum and exchange-correlation effects are properly taken into account in the density-density response function of the electrons, 
  differences with the data from the more advanced  simulations---e.g., molecular dynamics simulation of ions with the electrons treated by 
   density functional theory  (DFTMD)---will clearly indicate that non-linear screening effects might be of importance. 
   Therefore, with respect to the applicability to a quantum plasma with strongly coupled ions,  accurate analyses and checks of different existing theoretical models
   of the electronic density response (polarization) function
   are needed.  To this end, in this paper, 
   we study how the structural characteristics of the strongly coupled ions in a dense plasma are related to  the choice of the screened ion potential, in linear response. In particular, we are interested what is the effect of quantum degeneracy and electronic non-ideality on these potentials and, hence, on the ion structure.
  We are considering ion potentials that are obtained from ground state quantum Monte Carlo data (QMC), from  
    the random phase approximation (RPA), and various analytical models. 
 The particular focus is put on the use of the local field correction within the well-known Singwi-Tosi-Land-Sj\"olander approximation (STLS) \cite{stlsT0, stls}.  
  
  The motivation for the accurate and detailed investigation of the applicability of the  STLS based model for dense plasmas 
  and warm dense matter studies is that the STLS is conceptually clear, technically simple, and has been widely used in many applications where electronic correlations are important.
   For example, STLS-based methods were used to investigate  transport \cite{Bennadji, Reinholz95} and relaxation \cite{Benedict} properties,  stopping power ~\cite{Zwicknagel, Montanari,Gauthier, Wang, Barriga}, 
   the dynamical as well as the static structure factor \cite{Fortmann, Gregori, Plagemann, Saumon, RedmerIEEE}, 
   and thermodynamic properties \cite{BennadjiCPP, TanakaCPP, Scweng, Sjostrom} of dense plasmas  to mention but a few.
      Recently, considering the electron kinetic equation within a multiscale approach,  Graziani \textit{et al.} \cite{Graziani} developed an extended mean-field model which incorporates electronic correlations through the STLS ansatz.  
   Particularly, in the light of the latest developments in the fluid description of inhomogeneous quantum and non-ideal plasmas \cite{POP17, Hanno_1, Hanno_2} with the STLS closure relation, 
   the presented analysis of the applicability of the STLS description of the electronic correlations in the framework of the  multiscale approach is important and timely.
 
 The paper is structured as follows: In Sec.~ \ref{s:parameters}, the plasma state of interest and the corresponding dimensionless parameters are defined. 
 In Sec.~\ref{s:theory}, the theoretical formalism and the methods of calculations are presented. The results on  the structural properties of strongly coupled ions
 are shown in Sec.~ \ref{s:structure}. In the last section, we summarize our findings.


   \section{Plasma parameters}\label{s:parameters}
%
 In this paper we consider plasmas with degenerate electrons, i.e. the Fermi energy of the electrons, $E_F$, is larger or comparable to their characteristic thermal energy, $k_BT_e$, and the electronic degeneracy parameter obeys $\theta=k_BT_e/E_F\lesssim 1$. Furthermore, we consider plasmas with a high degree of ionization. This means,
 the temperature of the electrons is much higher than the atomic ionization energy. In dense plasmas, the ionization energy is drastically reduced due to quantum and correlation effects. 
 This effect 
 is referred to as the pressure ionization or Mott effect. 
 At $\theta\ll1$, the condition of pressure ionization 
 can be defined approximately by requiring the Fermi energy 
 to be greater than the modulus of the binding energy of the electron ($\sim e^2/2a_B$),  which yields for the Mott point, $r^M_s=a^M_e/a_B\approx 1.92$ \cite{Chabrier}, where  $a_e$ is the mean inter-electronic distance, and $a_B=\hbar^2/m_ee^2$ is the first Bohr radius, for the example of hydrogen.
 Indeed, this simple estimate gives the correct order of magnitude. First-principle path-integral Monte Carlo simulations indicate that 90\% of the bound states break up at a density of $r_s\sim 1.2$ \cite{Bonitz2005}. At higher temperature, the combined thermal and pressure ionization happens at lower density (larger $r_s$). Based on these estimates we will 
 consider, in the following, densities $n_e>10^{23}~{\rm cm^{-3}}$ and electronic temperatures $T_e\gtrsim 10^4~\rm K$.
  
  The electrons of the dense plasma are conveniently characterized by the dimensionless parameters $\theta$ and $r_s$ (we assume that the system is paramagnetic), whereas the state of the ions is determined by the 
  classical coupling parameter $\Gamma = Z^2e^2/(a k_BT_i)$, where $Ze$ is the ion charge, $T_i$ is the ion temperature, and $a=(4\pi n/3)^{-1/3}$ characterizes the mean distance between ions (Wigner–Seitz radius).
  In many experiments with dense plasmas, the ions are often  strongly coupled corresponding to a liquid state,
 $\Gamma>1$, but still far from the crystallization point \cite{Lyon, JeanCPP}. Therefore, we will consider ion coupling parameters in the range $1\leq \Gamma\leq 50$. 
  
  To have a complete physical picture of the non-isothermal plasma, in addition, the electron-ion coupling parameter, $\Gamma_{\rm ei}=Ze^2/ak_bT_e$, must be determined.
  This parameter indicates the validity of the theoretical method in use.  For example, when $\Gamma_{\rm ei}\ll1$ (an extremely dense plasma, $r_s\ll1$, or very hot electrons), 
  the ions can be described within the OCP. If  $\Gamma_{\rm ei}<1$, the ions still can be studied within the one-component plasma model, but with a screened ion-ion pair interaction potential \cite{Patrick}.
    In this case, the effect of electrons is absorbed into the effective ion potential. Moreover, due to the weak electron-ion coupling, the screening by the electrons can be described in the framework of linear response theory. 
    In the regime of strong coupling between the electronic and ionic subsystems, $\Gamma_{\rm ei}>1$, the creation of bound states (electron-ion recombination) must be considered. But this case is excluded by our choice of the density and temperature range, as explained above. 
    As a rule, in dense plasmas (warm dense matter) $\Gamma_{\rm ei}\lesssim 1$ \cite{JeanCPP, More}. 
 In Ref.~\cite{JeanCPP}, this was confirmed by orbital-free density functional theory. Furthermore, due to strong screening in dense plasmas, a more realistic electron-ion coupling parameter is $\Gamma_{\rm ei}\times \exp(-\kappa)$, 
 where  the dimensionless screening parameter (screening length in units of $a$) is in the range  $\kappa\sim 1-2$,  at the considered densities and temperatures [for a more accurate definition of the coupling parameter, see Ref.~\cite{Ott}].
 As a result, the actual electron-ion coupling is reduced even further. Also, for simplicity and without loss of generality, we set $Z=1$. 
  In this case, the ratio of the electron temperature to the temperature of ions can be related to $\theta$, $r_s$, and $\Gamma$ as 
  $T_e/T_{i}\simeq 1.84 \times (\theta/r_s)\, \Gamma$. For instance, at  $r_s=1.5$ and $\Gamma_i \lesssim 50$, 
  the degeneracy parameters $\theta=0.5$ and 0.1 imply a temperature ratio $T_e/T_{i} \lesssim 30$ and $T_e/T_{i} \lesssim 6$, respectively.
  
  A few relevant examples of experiments and associated plasma parameters are presented in Table~\ref{t:1}.
  There, densities, temperatures and the corresponding values of the degeneracy, density, and ionic coupling parameters are given. 
  These data confirm that the plasma parameters considered in this work are experimentally attainable and where the studied effects will be of relevance.
  Note that in many experiments, the plasma 
    undergoes a complex evolution during which the plasma parameters change substantially.
  For example, in Ref.~\cite{Strong_Coupling_and_Degeneracy_in_ICF}, 
  it was shown by analyzing experimental data \cite{ISI:A1997VY13200077, Paisner1994} that the ICF plasma enters the regime with strongly coupled ions ($\Gamma>1$) and quantum non-ideal electrons ($\theta<1$ and $r_s>1$).
  In the table we also list experiments where  non-isothermal (two-temperature) quantum plasmas with $\Gamma\gg1$ \cite{Ma, PhysRevLett.98.065002, PhysRevLett.102.115001} are observed.  For these experiments, the electron-ion temperature ratio, $T_e/T_i$,
  was evaluated in Refs.~\cite{Clerouin, Harbour, PhysRevE.92.013103}.
  
 \begin{widetext}

\begin{table} [t]
\centering
        \caption{Examples for experiments where two component quantum plasmas  with strongly coupled ions were realized}\label{t:1}
    \begin{tabular}{*5c}
        \hline
         \hline
         \\
        & \multicolumn{4}{c}{Plasma parameters} \\
       \cline{2-5}
       \\
    Systems and References & $n~[10^{23}~{\rm cm^{-3}}]$  & $T_e~ \& ~T_i~[10^3~{\rm K}]$ & $\theta~\&~r_s$ & $\Gamma$  \\    
         \hline
          \hline
          \\
       \vtop{ \hbox{\strut Cryogenic DT implosion on}\hbox{\strut OMEGA \cite{ISI:A1997VY13200077, Strong_Coupling_and_Degeneracy_in_ICF}}}    &   $2\lesssim n\lesssim 10$     &\vtop{\hbox{\strut $23\lesssim T_e\lesssim 230$ }\hbox{\strut $T_i=T_e$ }}   & 
       \vtop{\hbox{\strut $0.2\lesssim \theta \lesssim 0.8$  }\hbox{\strut $1.17\lesssim r_s \lesssim 2$ }}    &  $1\lesssim \Gamma \lesssim 6$  \\
         \hline
         \\
        \vtop{\hbox{\strut Direct-drive ignition at}\hbox{\strut the NIF \cite{Paisner1994, Strong_Coupling_and_Degeneracy_in_ICF}}}    &   $2.5\lesssim n\lesssim 3$     &\vtop{\hbox{\strut $69\lesssim T_e\lesssim 464$ }\hbox{\strut $T_i=T_e$ }}   & 
       \vtop{\hbox{\strut $0.2\lesssim \theta \lesssim 0.8$  }\hbox{\strut $1.75\lesssim r_s \lesssim 1.86$ }}    &  $0.5\lesssim  \Gamma \lesssim  3$  \\
        \hline
        \\
        \vtop{ \hbox{\strut Solid Be heated by 4-5 ${\rm keV}$}\hbox{\strut   pump photons \cite{ISI:000171262300031}}}    &   $2\lesssim n\lesssim 4$     &\vtop{\hbox{\strut $11\lesssim T_e\lesssim 110$ }\hbox{\strut $T_i=T_e$ }}   & 
       \vtop{\hbox{\strut $0.07\lesssim \theta \lesssim  1.15$  }\hbox{\strut $1.6\lesssim r_s\lesssim 2$ }}    & $2\lesssim  \Gamma \lesssim  10$  \\
         \hline
         \\
            \footnote{ $T_e/T_i$ was evaluated in Refs. \cite{Clerouin, Harbour}}  \vtop{\hbox{\strut Laser-driven shock-compressed }\hbox{\strut   aluminum \cite{Ma}}}    &   $n\simeq5.46$     &\vtop{\hbox{\strut $T_e\simeq 100$ }\hbox{\strut $T_e/T_i\simeq 5$ }}   & 
      \vtop{\hbox{\strut $\theta\simeq0.5$  }\hbox{\strut $r_s=1.435$ }}     &   $\Gamma \simeq 50$ \\
        \hline
        \\
        \footnote{$T_e/T_i$ was evaluated  in Ref. \cite{Harbour}}  \vtop{\hbox{\strut Laser-driven shock-compressed }\hbox{\strut  Be sample \cite{PhysRevLett.98.065002}}}    &   $n\simeq2.28$     &\vtop{\hbox{\strut $T_e\simeq 139$ }\hbox{\strut $2\lesssim T_e/T_i\lesssim  20$ }}   & 
       \vtop{\hbox{\strut $\theta\simeq0.88$  }\hbox{\strut $r_s=1.92$  }}    &   $7.5\lesssim \Gamma\lesssim 75$ \\
         \hline
         \\
           \footnote{$T_e/T_i$ was evaluated in Refs. \cite{PhysRevE.92.013103, Harbour}}            \vtop{\hbox{\strut Laser-driven shock-compressed }\hbox{\strut  Be sample \cite{PhysRevLett.102.115001}}}    &   $n\simeq6.7$     &\vtop{\hbox{\strut $T_e\simeq 150$ }\hbox{\strut $1.8\lesssim T_e/T_i\lesssim  6.5$ }}   & 
       \vtop{\hbox{\strut $\theta\simeq0.95$  }\hbox{\strut $r_s=1.34$  }}    &   $4\lesssim \Gamma\lesssim 16$ \\
         \hline
    \end{tabular}
\end{table}
 \end{widetext}
 
 \section{Theoretical description} \label{s:theory}

 \subsection{One component plasma model with effective ion-ion interaction}\label{s:model}
 
 First, we briefly discuss the route allowing to decouple the dynamics of the ions from that of the electrons, as well as the key assumptions of the model. 
 The ions are considered to be classical, and the electrons are treated fully quantum-mechanically in terms of continuous variables.
 Due to the large difference in masses, the electrons are assumed to adjust themselves instantaneously to a change in the ionic locations. 
 Therefore, a dense plasma is considered as a mixture of classical strongly coupled ions and a homogeneous quantum fluid of electrons. 
 These two systems are coupled through the 
 interaction energy,
 \begin{equation}\label{Uei}
  U_{\rm ei}=\frac{1}{2\Omega} \sum_{\vec k\neq0} \tilde \varphi_{\rm ei}(\vec k) \tilde{n}_i(\vec k)\tilde{n}_e(-\vec{k}),
 \end{equation}
 where $\tilde{n}$ denotes the deviation from the mean value of the density, and  $\tilde \varphi_{\rm ei}(\vec k)$ is the Fourier transform of the bare electron-ion interaction potential.
As long as the potential energy  $U_{\rm ei}$ is smaller than the quantum kinetic energy of the electrons (which, at $\theta\ll1$, is of the order of $E_F$), 
we may treat ion-electron interaction as a perturbation. 
For instance, in the lowest order approximation, $U_{\rm ei}=U_{\rm ei}^{(1)}=0$, we recover the OCP model for the ions, meaning the electrons do not respond at all to the field of the ions.    
In the second order approximation,  the response of the electrons is linear in the perturbing field of the ions, $\tilde n_e(-\vec k)=\chi_e(\vec k)\tilde \varphi_{\rm ei}^{*}(\vec k) \tilde n_{i}(-\vec k)$, where $\chi_e(\vec k)$ is the 
 static electron-density response function of a translationally invariant  electron system. Then we find
  \begin{equation}\label{Uei_1}
  U_{\rm ei}^{(2)}=\frac{1}{2\Omega} \sum_{\vec k\neq0} \left| \tilde \varphi_{\rm ei}(\vec k)\right|^2\chi_e(\vec k) \tilde{n}_i(\vec k)\tilde{n}_i(-\vec{k}).
 \end{equation}
From Eq.~(\ref{Uei_1}) we see that $U_{\rm ei}^{(2)}$ depends only on the variables of the ions. 
Therefore, defining the screened ion-ion interaction potential as \cite{Hansen}
\begin{multline}\label{eff_pot_1}
 \Phi(\vec r_{j^{\prime}},\vec r_j)= \frac{Z^2e^2}{|\vec r_{j^{\prime}}-\vec r_j|}+\\ \int\! \frac{\mathrm{d}^3k}{(2 \pi)^3 } \left| \tilde \varphi_{\rm ei}(\vec k)\right|^2  \chi_e(\vec k)  \,\, e^{i \vec k \cdot (\vec r_{j^{\prime}}-\vec r_j)} \quad,
\end{multline}
the decoupled total Hamiltonian of the system can be written as
\begin{equation}\label{Hamiltonian}
 H=H_{i}\left(\vec R, \vec P\right)+H_{e}[n_e],
\end{equation}
where $\vec R={\vec r_1,...,\vec r_{N_i}}$ and $\vec P={\vec p_1,...,\vec p_{N_i}}$ is the complete set of ionic coordinates and momenta.
The Hamiltonian of the ionic subsystem reads,
\begin{equation}
 H_{i}\left(\vec R, \vec P\right)=K_i\left(\vec P\right)+\sum_{j=1}^{N_i}\sum_{j^{\prime}>j}^{N_i}\Phi(\vec r_{j^{\prime}},\vec r_j),
\end{equation}
 with $K$ being the kinetic energy of the ions. Further,  $H_e[n_e]$ in Eq.~(\ref{Hamiltonian}) is the Hamiltonian of the electronic reference system \textit{without perturbation by the field of the ions}. 
The hamiltonian (\ref{Hamiltonian}) can  then be used for the description of various properties of the system \cite{Zubarev}.  

In the following, we consider the system without the influence of external fields (electric or magnetic) and take the homogenous electron gas as the reference system.
Taking  the bare electron-ion interaction potential in the form of the Coulomb potential,  $\tilde \varphi_{\rm ei}(\vec k)=4\pi Ze^2/k^2$, 
 one readily recovers the widely used expression for the screened potential from Eq.~(\ref{eff_pot_1})
\begin{equation} \label{POT_stat}
\Phi(\vec r)   = \int\!\frac{\mathrm{d}^3k}{2 \pi^2 } \frac{Q^2}{k^2 \epsilon(\vec k, \omega=0)} \,\, e^{i \vec k \cdot \vec r} \quad,
\end{equation}
with $\epsilon^{-1}(\vec{k},\omega)$ being the inverse dielectric response function of the electrons 
\begin{equation}\label{dynamic_epsilon_0}
 \epsilon^{-1}(\mathbf{k},\omega) = 1 +\frac{4\pi e^2}{k^2}\chi_e(\mathbf{k},\omega).
\end{equation}
It  should be noted that, if for $\varphi_{\rm ei}$ instead of the Coulomb potential an electron-ion pseudopotential (such as the so-called empty core potential) is chosen, the effective potential (\ref{eff_pot_1}) must be used.

Further, all electronic correlation effects are conviently incorporated in the so-called local field correction $G$ that enters the density response function via
\begin{equation}\label{chi_G}
\chi_e^{-1}(\vec k, \omega)=\chi_0^{-1}(\vec k, \omega)+\frac{4\pi e^2}{k^2}\left[G(\vec k, \omega)-1\right],
\end{equation}
where $\chi_0$ denotes the finite temperature ideal density response function of the electron gas~\cite{quantum_theory}.


A highly successful way to approximately determine the static local field correction is provided by the self-consistent static STLS scheme~\cite{stlsT0,stls}, which is based on the relation
\begin{equation}
G^\textnormal{STLS}(\mathbf{k},0) = -\frac{1}{n} \int\frac{\textnormal{d}\mathbf{k}^\prime}{(2\pi)^3}
\frac{\mathbf{k}\cdot\mathbf{k}^\prime}{k^{\prime 2}} [S^{\text{STLS}}(\mathbf{k}-\mathbf{k}^\prime)-1]\;,
\label{G_stls}
\end{equation}
where the static structure factor $S^{\textnormal{STLS}}$ is calculated according to the fluctuation-dissipation theorem as
\begin{equation}\label{flucDis}
 S^\text{STLS}(\mathbf{k}) = -\frac{1}{\beta n}\sum_{l=-\infty}^{\infty} \frac{k^2}{4\pi e^2}\left(\frac{1}{\epsilon(\mathbf{k},z_l)}-1\right),
\end{equation}
where the summation is over the Matsubara frequencies, $z_l=2\pi il/\beta\hbar$. 
The inverse dielectric function is computed via Eq.~(\ref{dynamic_epsilon_0}) using 
$\chi_0$ and $G^\text{STLS}$. 
Thus, Eqs.~(\ref{dynamic_epsilon_0})
--(\ref{flucDis}) form a closed set of equations that can be solved self-consistently, 
to yield the static dielectric function $\epsilon^\text{STLS}(\mathbf{k},0)$. In the following, the potential computed in this way will be referred to as the \textit{STLS screened potential}, whereas the \textit{RPA screened ion potential} corresponds to the case $G=0$. 

The most reliable data for the static local field correction are those obtained from \textit{ab-initio} quantum Monte Carlo simulations.
 Corradini \textit{et al.} \cite{Corradini} provided an accurate parametrization of the ground state quantum Monte Carlo data \cite{Moroni} for $r_s=2,~5,~10$ and 100.
 In order to check features of the radial pair distribution function and static structure factor of the strongly coupled ions interacting via 
 the STLS screened ion potential, we provide comparisons---in the context of the considered parameters---with results computed using the static local field correction 
 of Ref.~\cite{Corradini}  at $r_s=2$ and $\theta=0.01$ (referred to as the \textit{QMC data based potential}, below).

Using in Eq.~(\ref{POT_stat}) the first order result of the long wavelength expansion of the inverse 
 ideal (RPA) response function, i.e. $\chi_e^{-1}(k)\approx\chi_0^{-1}(k\to0)\approx \tilde a_0$, the widely used Yukawa-type  screened ion potential is obtained,
\begin{equation}\label{Yukawa}
\Phi_{Y}(r;n,T)=\frac{Q^2}{r}\, e^{-k_{s}r},
\end{equation}
where $\tilde a_0=4\pi e^2/k_{Y}^2$, and the finite-temperature inverse screening length, $k_s$, is equal to $k_{Y}(n,T)=\left[\frac{1}{2} k_{TF}^2 \theta ^{1/2} I_{-1/2}(\beta \mu)\right]^{1/2}$,  with the inverse electron temperature $\beta=1/k_BT_e$, and $\beta \mu$ is determined by the normalization, $n= \sqrt{2} I_{1/2}(\beta \mu) / \pi^2\beta^{3/2}$.
It is worth noting that $k_{Y}$ corresponds to the Thomas-Fermi and Debye-H\"uckel expressions 
in the fully degenerate  and classical limits, respectively.
The Thomas-Fermi wave number is given by $k_{TF}=\sqrt{3}\omega_{p}/v_{F}$, and $I_{-1/2}$ is the Fermi integral of order $-1/2$. 
In the context of dense plasmas, Eq.~(\ref{Yukawa}) is often referred to as the Thomas-Fermi potential (TF).

Substituting into  Eq.~(\ref{POT_stat})  the second order result of the long wavelength expansion of the inverse RPA response function, 
  i.e. $\chi_e^{-1}(k)=\chi_0^{-1}(k\to0)\approx \tilde a_0+\tilde a_2\cdot k^2$ \cite{POP2015}, one recovers the analytical model by Stanton and Murillo (SM)~\cite{Murillo}
\begin{align}\label{eq:sm<}
\phi(r;n,T) = \frac{Q^2}{2r}\left[(1+b)\, e^{-k_+ r} + (1-b)\, e^{-k_- r}\right],
\end{align}
where $b = 1/\sqrt{1 - \alpha^{SM}}$ and $k_\pm = k_{TF}(1\mp \sqrt{1-\alpha^{SM}})^{1/2} / \sqrt{\alpha^{SM}/2}$,  $\alpha^{SM} = 3 \sqrt{8 \beta} \lambda {I'}\!\!_{-1/2}(\beta \mu) / \pi$, $\lambda = 1/9$,  
and $I'_p(\beta \mu)$ is the derivative of the Fermi integral with respect to $\beta \mu$. Further, the inverse Thomas-Fermi screening length at finite temperature is given by $k_{TF} = (4I_{-1/2}(\eta_0)/\pi\sqrt{2\beta})^{1/2}$.
  In Ref.~\cite{Murillo}, at $\alpha^{SM}>1$ the SM potential (\ref{eq:sm<}) was expressed in a somewhat different form to show the appearance of the oscillatory pattern in a certain range of densities and temperatures.
  The expression for this case can be easily found from Eq.~(\ref{eq:sm<}) using Euler's formula relating trigonometric functions and the complex exponential function.    
 Therefore,  the SM potential in the form of Eq.~(\ref{eq:sm<}) can be used regardless of the value of $\alpha^{SM}$.
  We note that the potential (\ref{eq:sm<}) was originally derived 
  on the basis of the  Thomas-Fermi model with the first order gradient correction to the non-interacting free energy density functional. 
 For the ground state ($\theta=0$) the potential~(\ref{eq:sm<}) was derived by Akbari-Moghanjoughi \cite{Akbari} using linearized quantum hydrodynamic equations.
 For a detailed discussion of the mutual connection between DFT, the quantum hydrodynamics model, and linear response theory we refer the reader to Ref.~\cite{POP17}. 
  
  The potentials (\ref{Yukawa}) and (\ref{eq:sm<}) correctly describe the screening of the ion potential at large distances but neglect non-ideality (correlation) effects.
 It is important to stress that potentials (\ref{Yukawa}) and (\ref{eq:sm<}) are lower order approximations with respect to the full (nonlocal) RPA description (see Ref.~\cite{POP2015}). 

 We use the STLS screened potential, the RPA screened potential, the Yukawa potential, the SM potential, and the QMC data based potential for the 
 calculation of the radial pair distribution function  and static structure factor of strongly coupled ions. This is done by implementing the aforementioned potentials into the solution of the  
  Ornstein-Zernike integral equation in the hypernetted chain approximation (HNC). This allows us to perform a detailed analysis of the applicability limits of the different potentials 
 and to identify the effects related to the electronic correlations as well as non-locality. As a cross-check, the main conclusions obtained using HNC are tested by molecular dynamics simulation (MD). 
 The comparison of the STLS screened potential with other potentials considered in this work was recently presented by Moldabekov \textit{et al.} \cite{CPP2017}.
  
  \begin{figure}[h]
\includegraphics[width=0.4\textwidth]{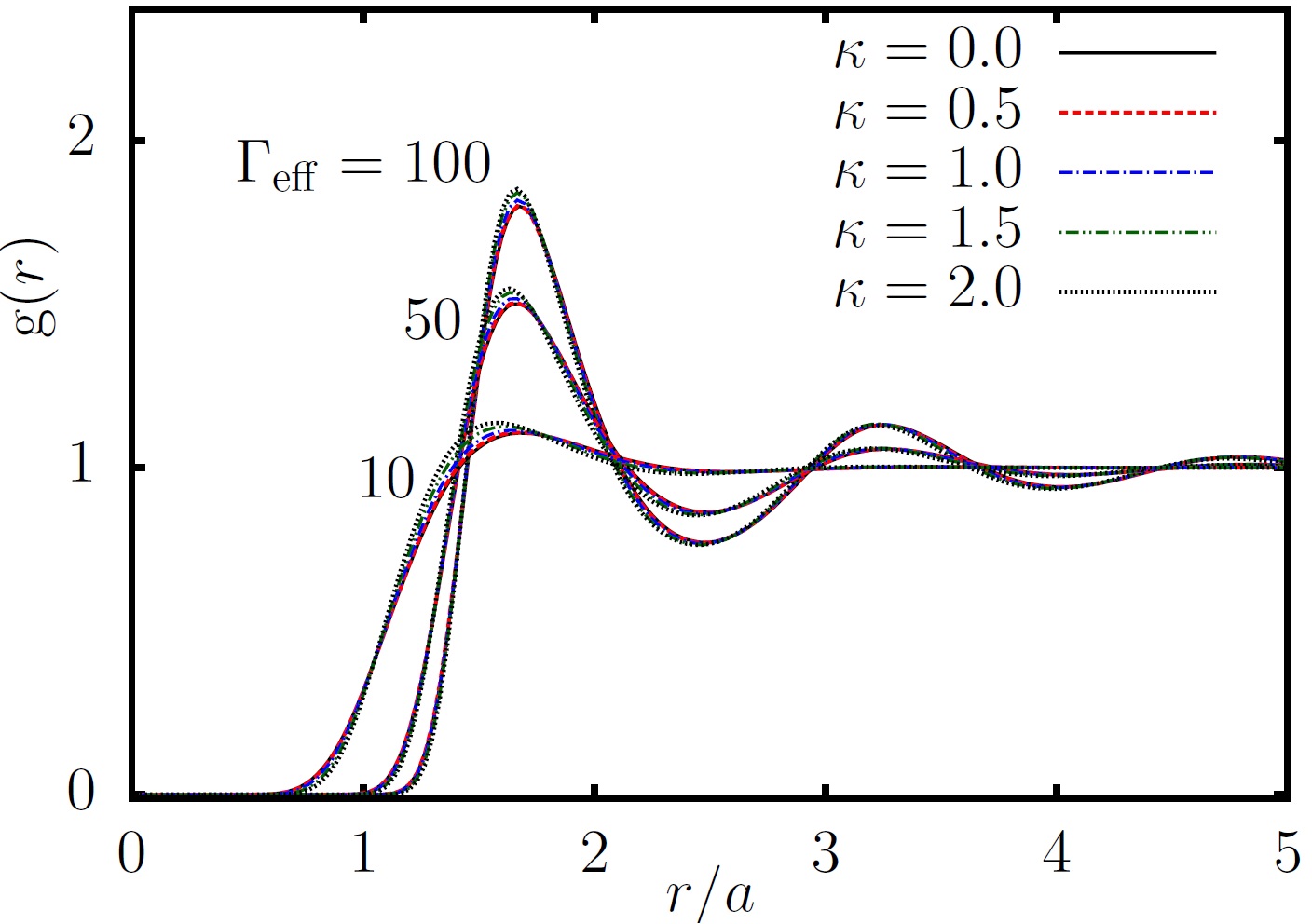}
\caption{The HNC result for the radial pair distribution function calculated for different screening parameters $\kappa$ and effective coupling parameters $\Gamma_{\rm eff}$.}
\label{fig:Geff_HNC}
\end{figure}

  \begin{figure}[h]
\includegraphics[width=0.4\textwidth]{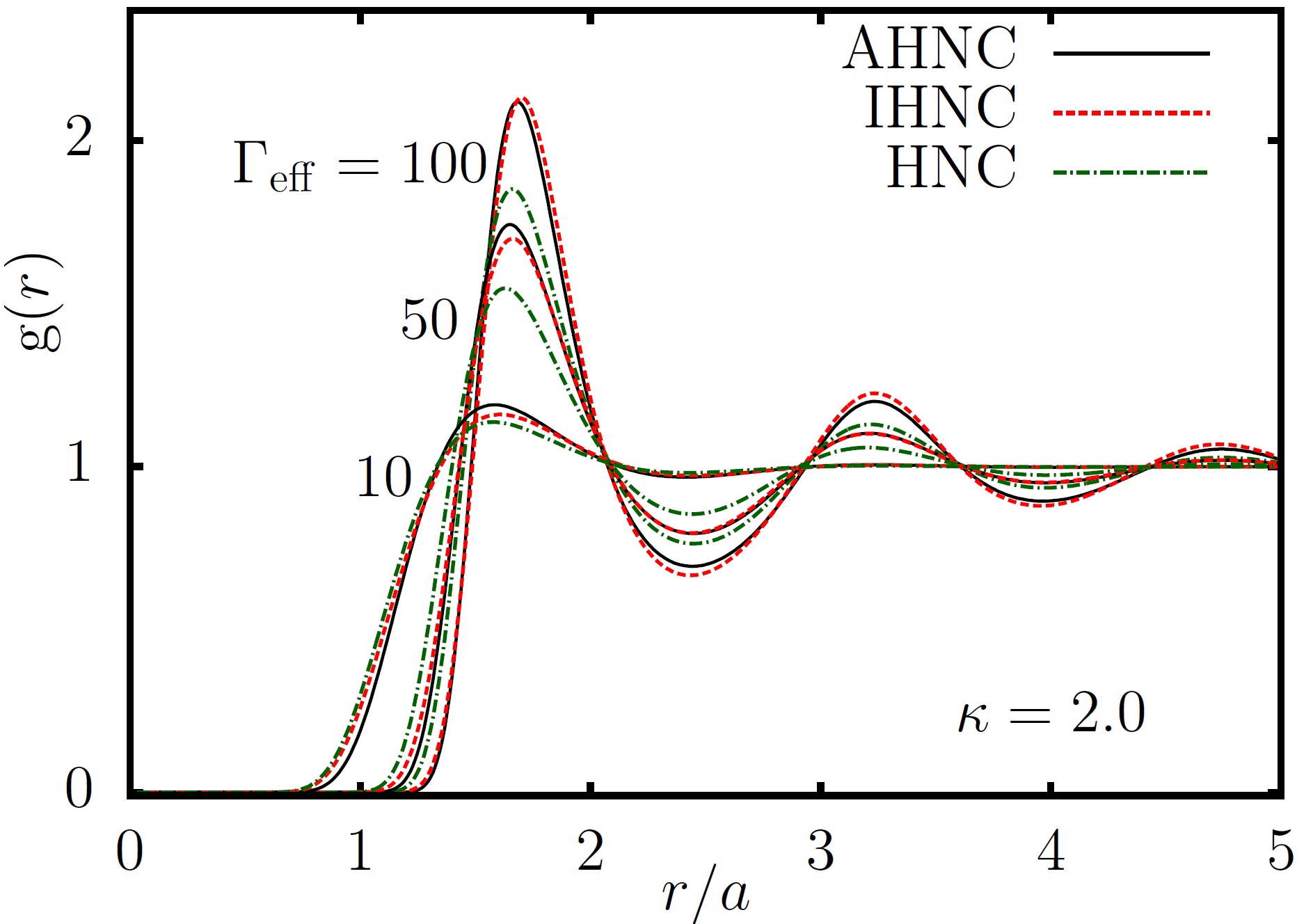}
\caption{Radial pair distribution function calculated for different effective coupling parameters $\Gamma_{\rm eff}$ using the HNC approximation, the AHNC (with bridge function by Ng \cite{Ng}), and the IHNC (with bridge function by
 Daughton \textit{et al.} \cite{Daughton}).}
\label{fig:Geff_comp}
\end{figure}

  \begin{figure}[h]
\includegraphics[width=0.4\textwidth]{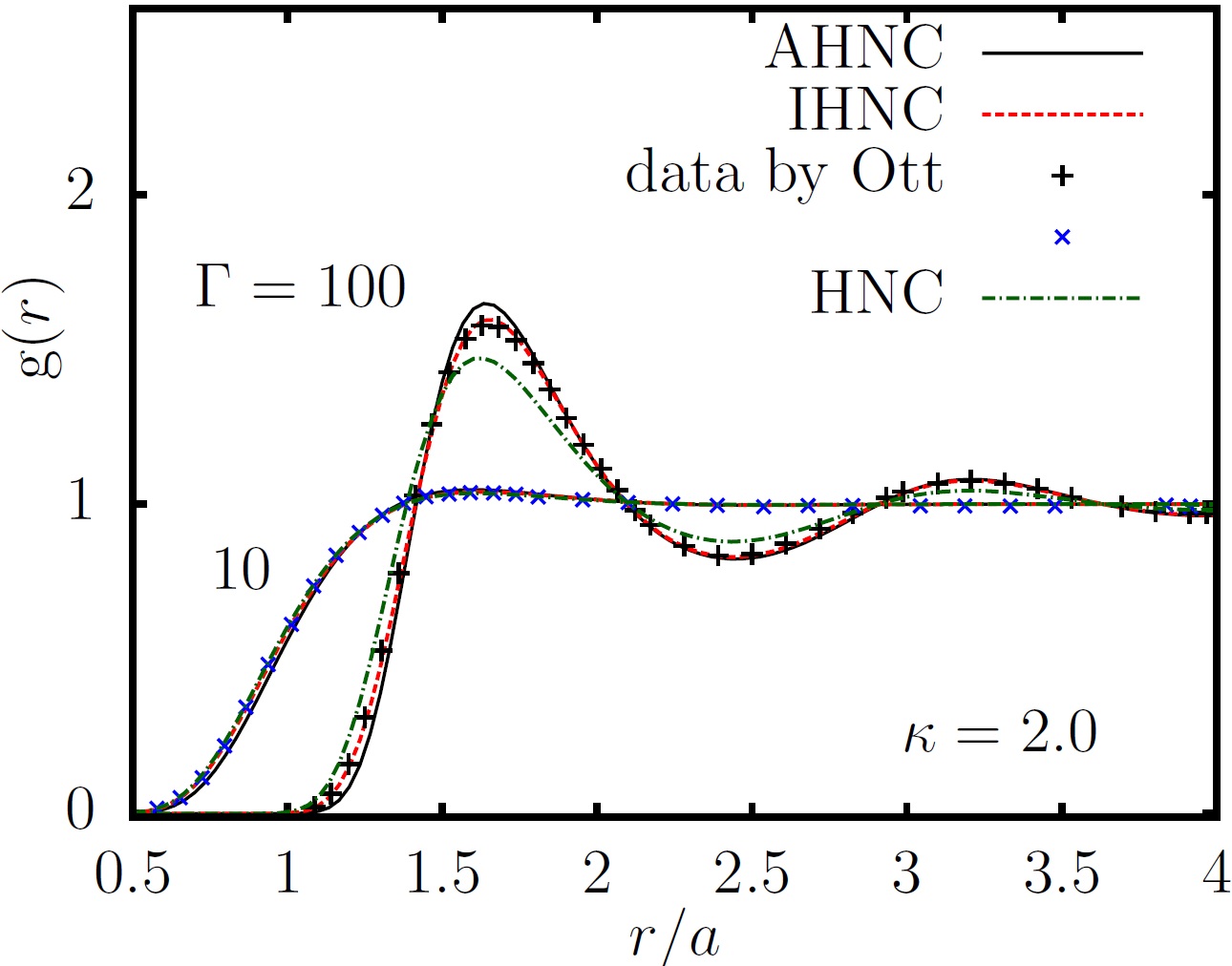}
\caption{The same as in Fig.~\ref{fig:Geff_comp} but here comparison with the MD results by Ott \textit{et al.}~\cite{Ott} at $\Gamma=10,~100$, and $\kappa=2.0$.}
\label{fig:grYOCP}
\end{figure}

 It is worth noting that, at $\theta\lesssim 1$, previous DFTMD results suffered from a lack of reliable input for the exchange-correlation free energy density of electrons. 
 In fact, the first accurate results on the exchange-correlation free energy density on the level of the local density approximation have been obtained only recently \cite{groth_prl, Tobias, dornheim_pr_2018}. 
  Additionally, most of the DFTMD simulations were done for isothermal plasmas (see discussions in Ref.~\cite{Clerouin}). 
  Another method for the description of two-component dense plasmas is based 
   on the theory of quantum interaction potentials between plasma particles \cite{Filinov, Wunsch}. 
   At the considered plasma parameters, this method suffers 
   from an approximate treatment of electronic quantum effects and, more importantly, from the uncertainty in the choice of the so-called electron-ion temperature, 
   which is needed for the determination of the electron-ion quantum potential \cite{Seuferling, Bredow, Shaffer}. 
    Therefore, the present approach of treating the ions in the OCP framework  with an effective pair interaction potential, obtained from a linear response description of the electrons,
     is highly valuable for the understanding of dense plasmas out of equilibrium \cite{Patrick}.

\subsection{The HNC approximation in terms of an effective coupling parameter $\Gamma_{\rm eff}$}\label{s:HNC}

The Ornstein-Zernike relation is given by
 \begin{equation}\label{eq:OZ}
  h(r)=c(r)+n\int c(r^{\prime}) h(|\vec r-\vec r^{\prime}|) \mathrm{d}\vec{r}^{\prime}\, ,
 \end{equation}
 and the formally exact closure reads
 \begin{equation}\label{eq:closure}
  g(r)=\exp {\left[ -\beta u(r) +h(r)-c(r) +B(r)\right]},
 \end{equation}
where $\beta=(k_BT)^{-1}$, $u(r)$ is the pair interaction potential, $g(r)$ the radial pair distribution function (RPDF), $h(r)=g(r)-1$ the total correlation function,
 $c(r)$ the direct correlation function, and $B(r)$ is the bridge function.
 The hypernetted chain approximation corresponds to the case $B(r)=0$. Details about the numerical solution of Eqs. (\ref{eq:OZ}) and (\ref{eq:closure}) are given in Appendix A.
 
 It is well known that the RPDF of the OCP and of the YOCP can be unified by introducing an effective coupling parameter, $\Gamma_{\rm eff}$, that  characterizes the strength of correlations determined from the shape of the RPDF.
 Based on accurate MD simulation data, Ott \textit{et al.} \cite{Ott} found the following simple parametrization of  $\Gamma_{\rm eff}$,
 \begin{align}
  \Gamma_{\rm eff}(\Gamma, \kappa)&= f(\kappa)\cdot \Gamma,\\
  f(\kappa)&=1-0.309 \kappa^2+0.08 \kappa^3,
 \end{align}
in the range $0\leq \kappa\leq 2$, and $1\leq \Gamma_{\rm eff}\leq 150$, where $\kappa=k_s a$ is the screening parameter.

We use this approach to test the performance of the HNC for an accurate computation of the RPDF. 
 We find that the HNC results nicely follow a one-to-one mapping between the RPDF of Coulomb and Yukawa systems, similar to the MD results \cite{Ott}. 
This is illustrated in Fig.~\ref{fig:Geff_HNC}. Secondly, it is revealed that for the accurate description of the RPDF the HNC can be used up to $\Gamma_{\rm eff}\simeq 10$. 
To illustrate this, in Fig.~\ref{fig:Geff_comp}  the comparison of the HNC results with the RPDF calculated using the bridge function by Daughton \textit{et al.} \cite{Daughton}, for the Yukawa system, 
and  by Ng \cite{Ng}, for the Coulomb system, are shown. Following the notation of Ref.~\cite{Bruhn}, the use of the bridge function by Daughton \textit{et al.} is denoted as the IHNC (``improved HNC'') and the bridge function by Ng is denoted as AHNC (``adjusted HNC'').  The IHNC gives very good agreement with the MD data \cite{Daughton} and, thereby, can be considered as the ``exact'' one. 
 It is worth noting that, even at $\kappa=2$, the AHNC yields good agreement with the IHNC results. This is one more illustration of a certain level of
  universality of the bridge functions as it was suggested by Rosenfeld and Ashcroft \cite{Rosenfeld}. 
  
  The comparison with the MD data by Ott \textit{et al.} \cite{Ott, Bruhn} is given in Fig.~\ref{fig:grYOCP} 
  and confirms the correctness of our numerical solution of the  Ornstein-Zernike  equation and allows to quantify the accuracy of the different approximations.

As an example of our findings, we note  that at $\kappa=2$ the condition  $\Gamma_{\rm eff}=10$ gives $\Gamma=25$ as the maximal value of the coupling parameter up to which a good agreement between the HNC and MD data on the RPDF 
 is observed. This indicates that at stronger screening, $\kappa>2$, the agreement between the HNC data and the MD result can be extended well beyond $\Gamma=25$. Indeed, as it is shown below, this is confirmed 
  by the calculations based on the STLS screened potential, where screening is significantly stronger in comparison with the Yukawa potential (\ref{Yukawa}) due to  electronic non-ideality. 
  
  The screened pair interaction potentials used in this work depend on the temperature and density of the system and, therefore,  belong to the class of state-dependent potentials.
 Many of the widely used integral equations, like the Ornstein-Zernike relation, were originally derived considering a pair interaction which does not have a dependence on the  temperature and density (e.g. Coulomb potential).
 Therefore,  the use of the solution of these integral equations on the basis of the screened pair interaction potentials must be done with caution (see discussions in Refs.~ \cite{Tejero, Adamo}).
 In this work, the main results were verified by MD simulations of ions interacting through a screened ion potential.   
        
  \begin{figure}[h]
\includegraphics[width=0.45\textwidth]{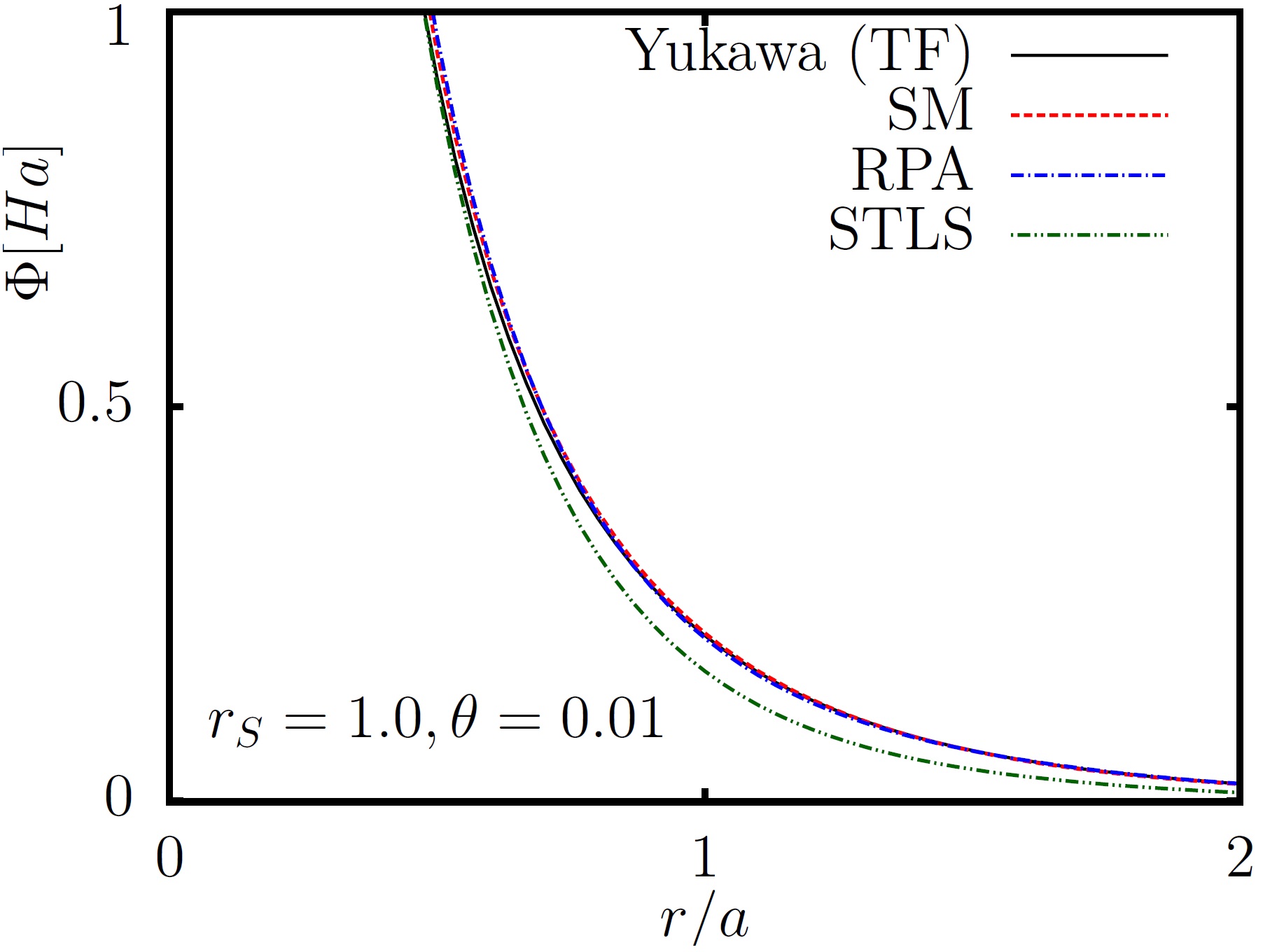}
\caption{Screened ion potential in different approximations.}
\label{fig:potrs1}
\end{figure}
  \begin{figure}[h]
\includegraphics[width=0.48\textwidth]{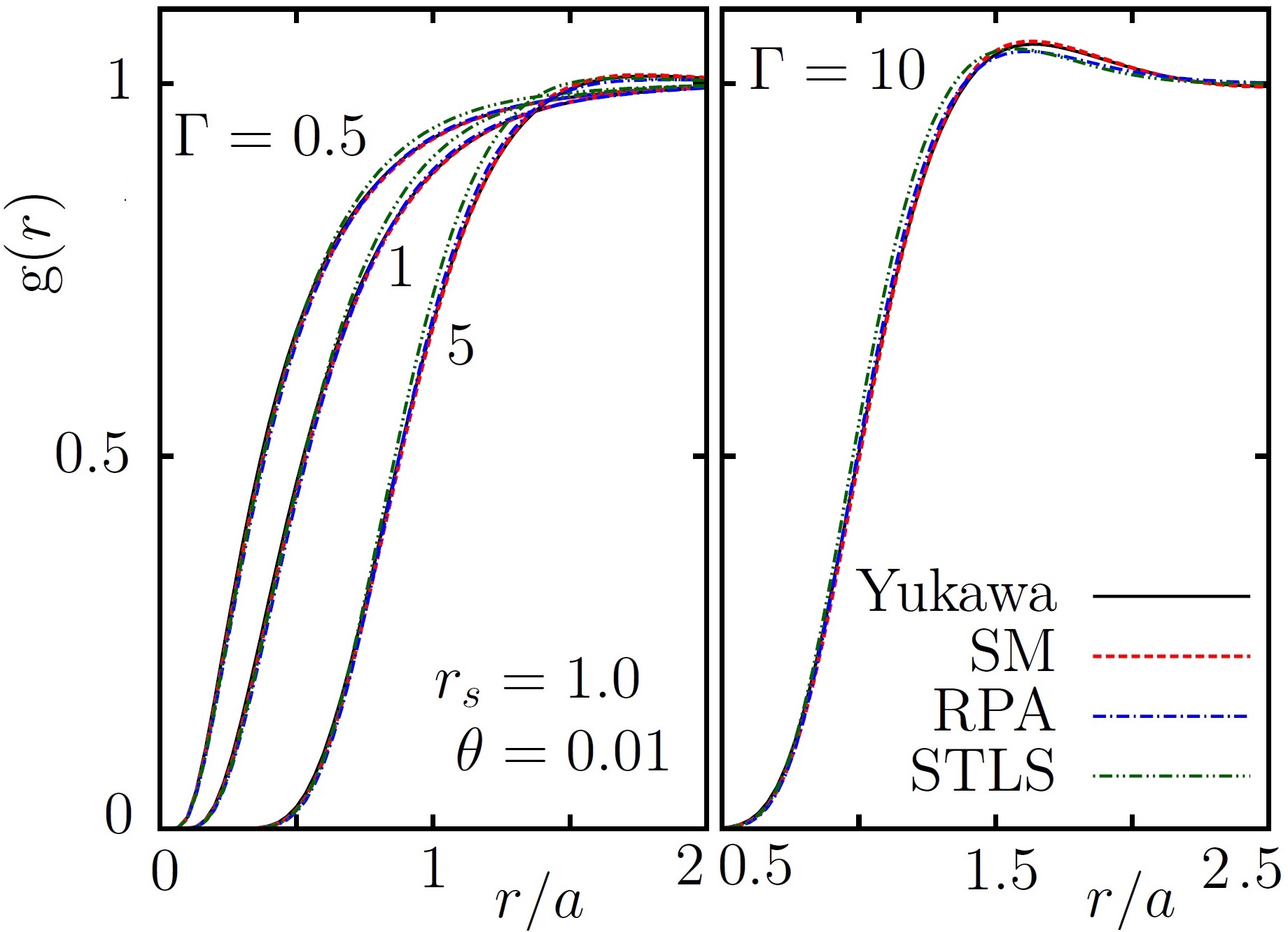}
\caption{Radial pair distribution function calculated using different screened ion potentials with $\Gamma=0.5,~1,~5$ and 10.}
\label{fig:rs1theta0v2}
\end{figure}

  \begin{figure}[h]
\includegraphics[width=0.48\textwidth]{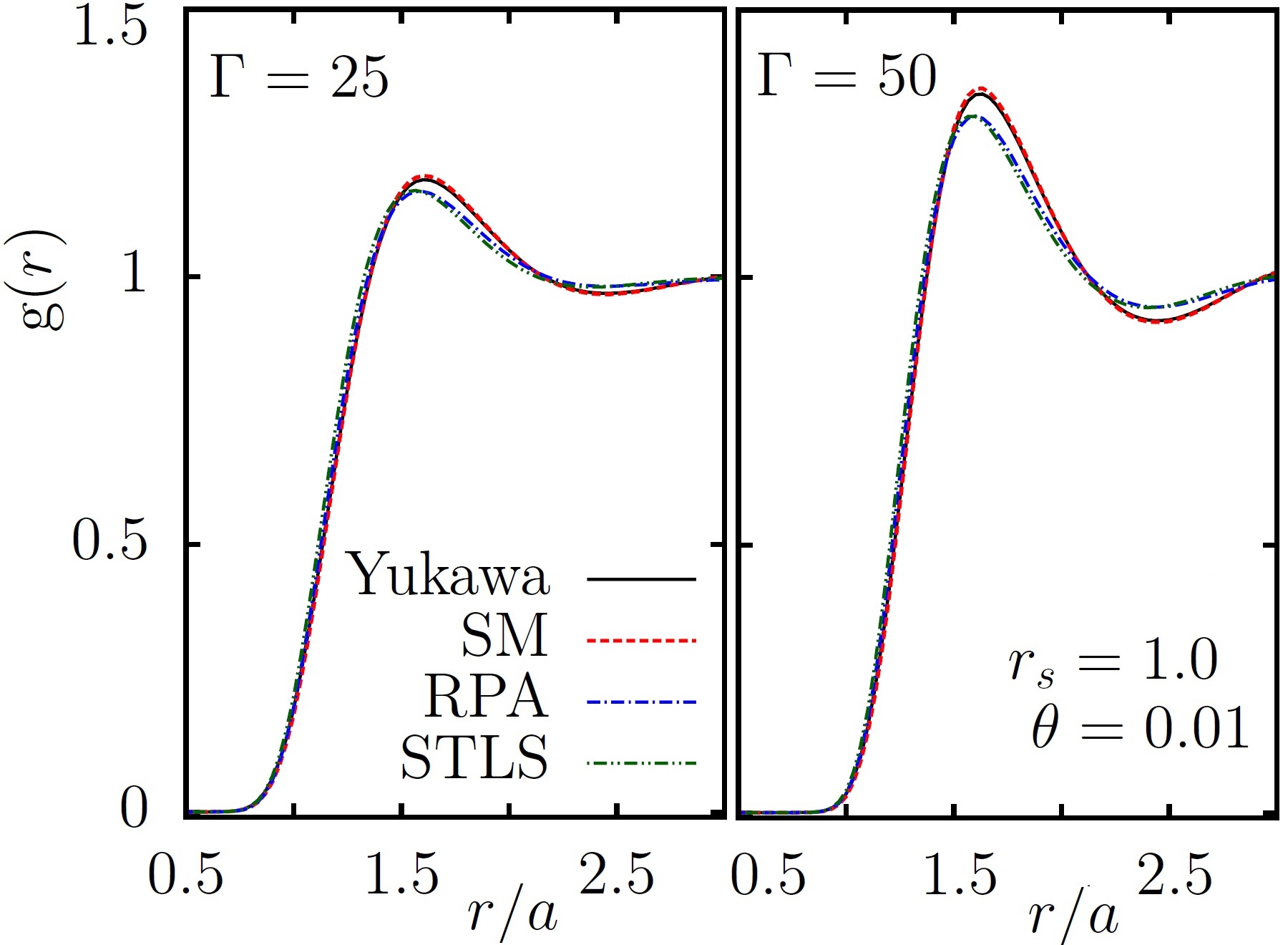}
\caption{The same as in Fig.~\ref{fig:Geff_comp} but for $\Gamma=25$ and 50.}
\label{fig:rs1theta0G25-50v2}
\end{figure}

  \begin{figure}[h]
\includegraphics[width=0.45\textwidth]{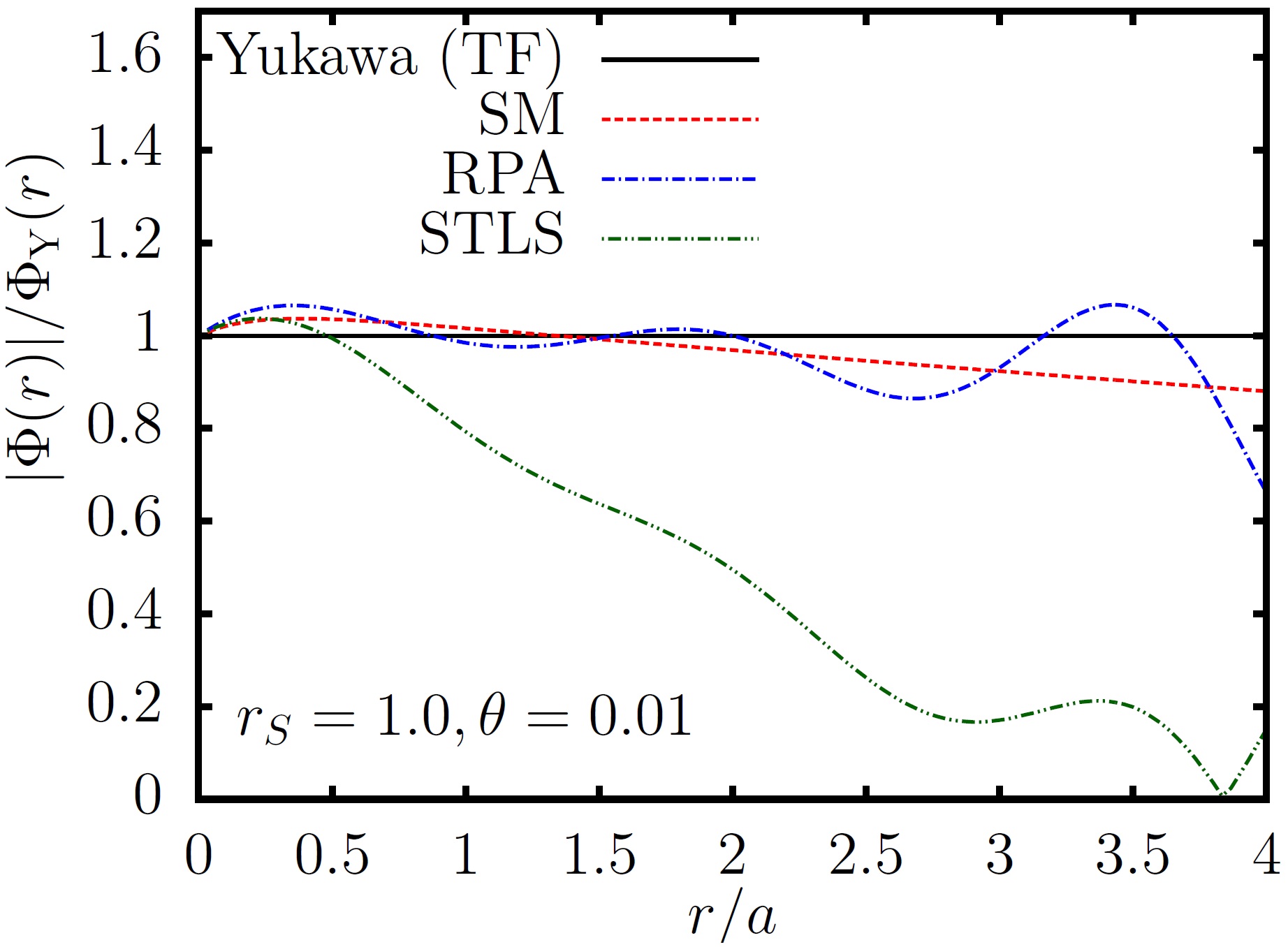}
\caption{The ratio of the different screened potentials to the Yukawa potential~(\ref{Yukawa}).}
\label{fig:mod_pot}
\end{figure}

  \begin{figure}[h]
\includegraphics[width=0.43\textwidth]{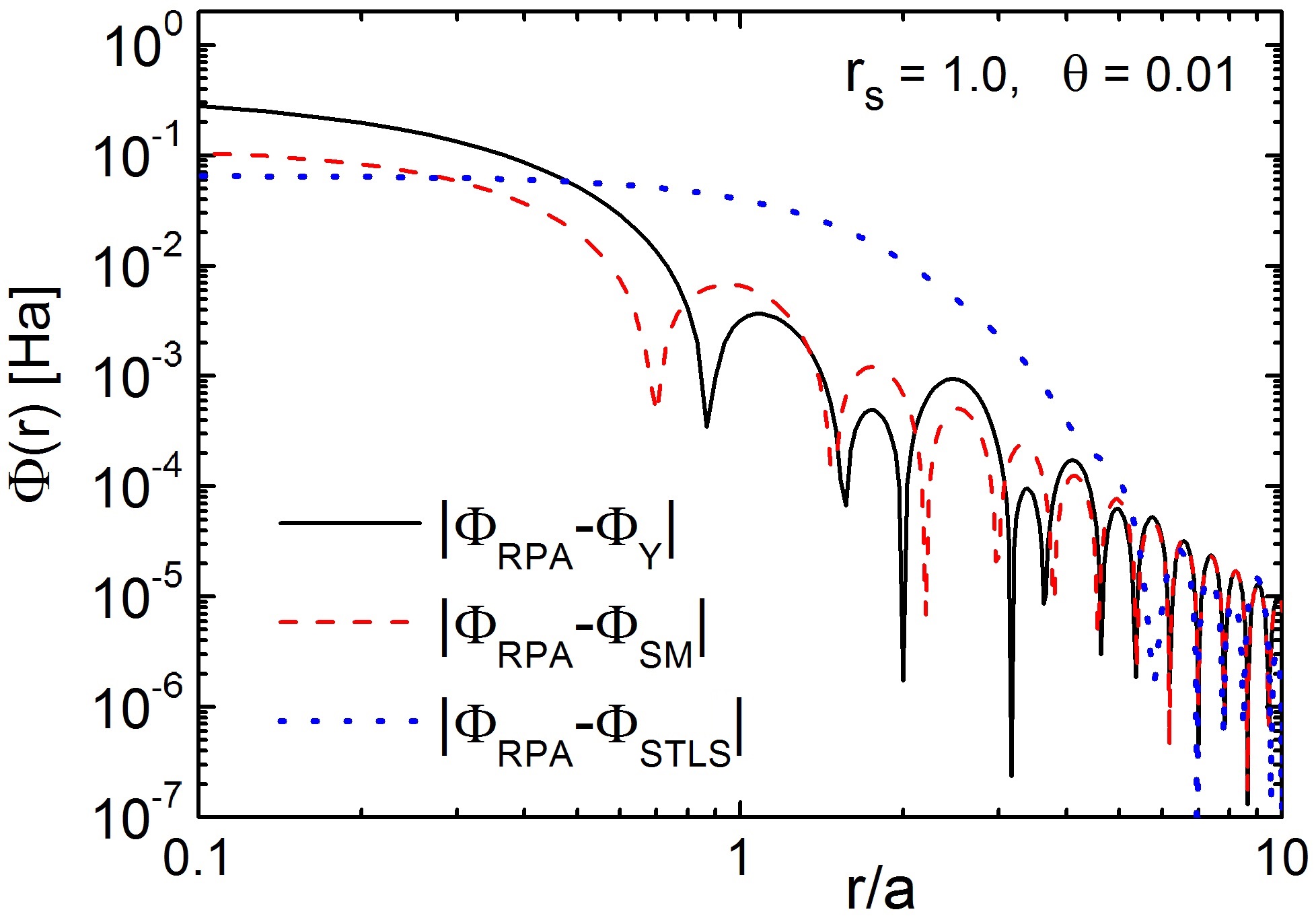}
\caption{The absolute value of the difference between the RPA screened potential and other potentials.}
\label{fig:dif_pot2}
\end{figure}

  \begin{figure}[h]
\includegraphics[width=0.45\textwidth]{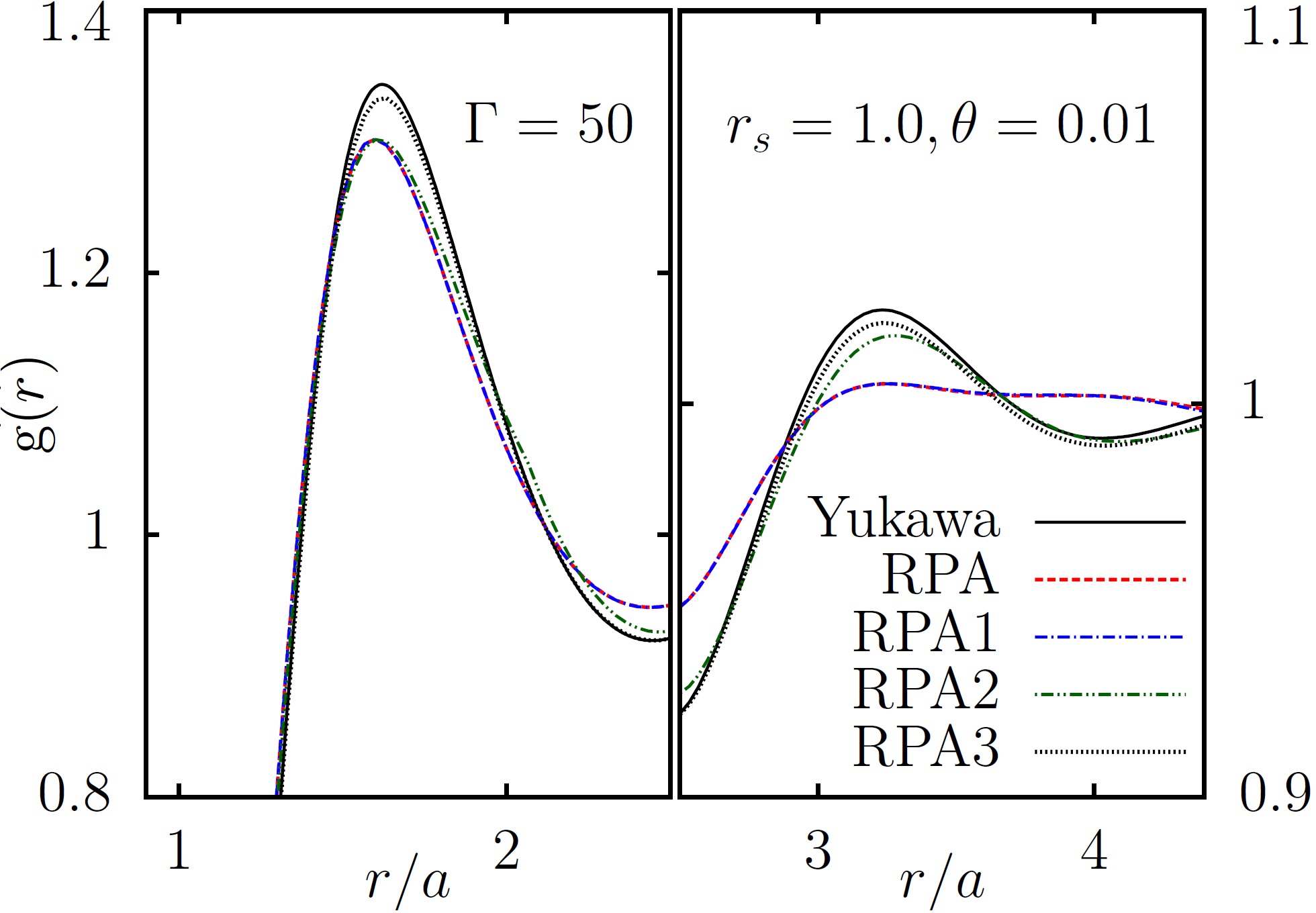}
\caption{Radial pair distribution function calculated using the Yukawa potential~(\ref{Yukawa}), the RPA screened potential, and the test potential (\ref{hybrid}).}
\label{fig:gr_osc50}
\end{figure}

    \begin{figure}
 \includegraphics[width=0.45\textwidth]{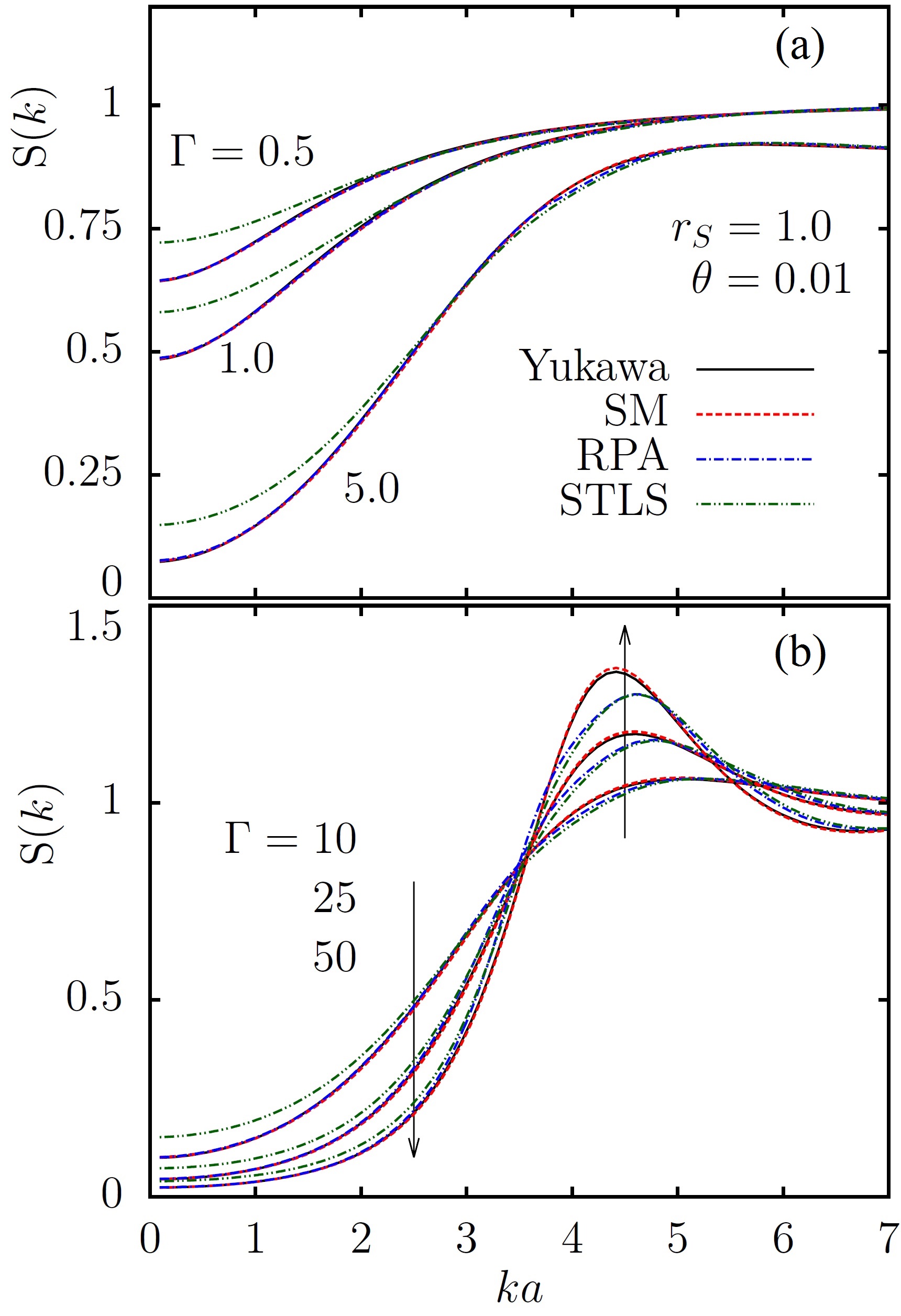}
 \caption{Static structure factor computed using different screened potentials. In the upper figure (a) the curves for $\Gamma=5.0$ are shifted relatively to the cases $\Gamma=0.5$ and 1.0 for the better visibility.}
 \label{fig:rs1Sk}
 \end{figure}
 
   \begin{figure}
\includegraphics[width=0.5\textwidth]{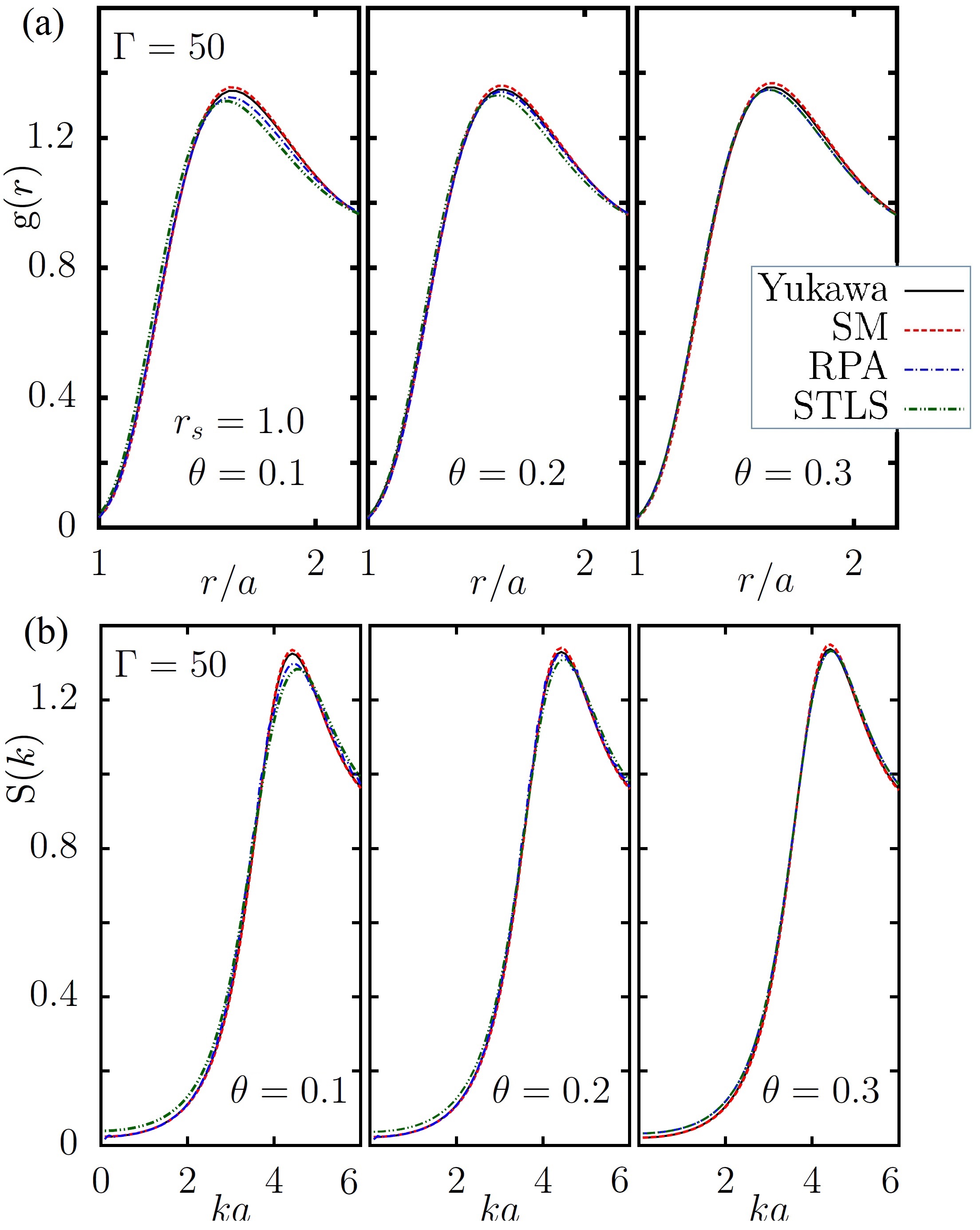}
\caption{Radial pair distribution function and structure factor computed at different values of the degeneracy parameter $\theta$ of electrons at $r_s=1.0$ and $\Gamma=50$.}
\label{fig:gr_theta}
\end{figure}

  \begin{figure}
\includegraphics[width=0.38\textwidth]{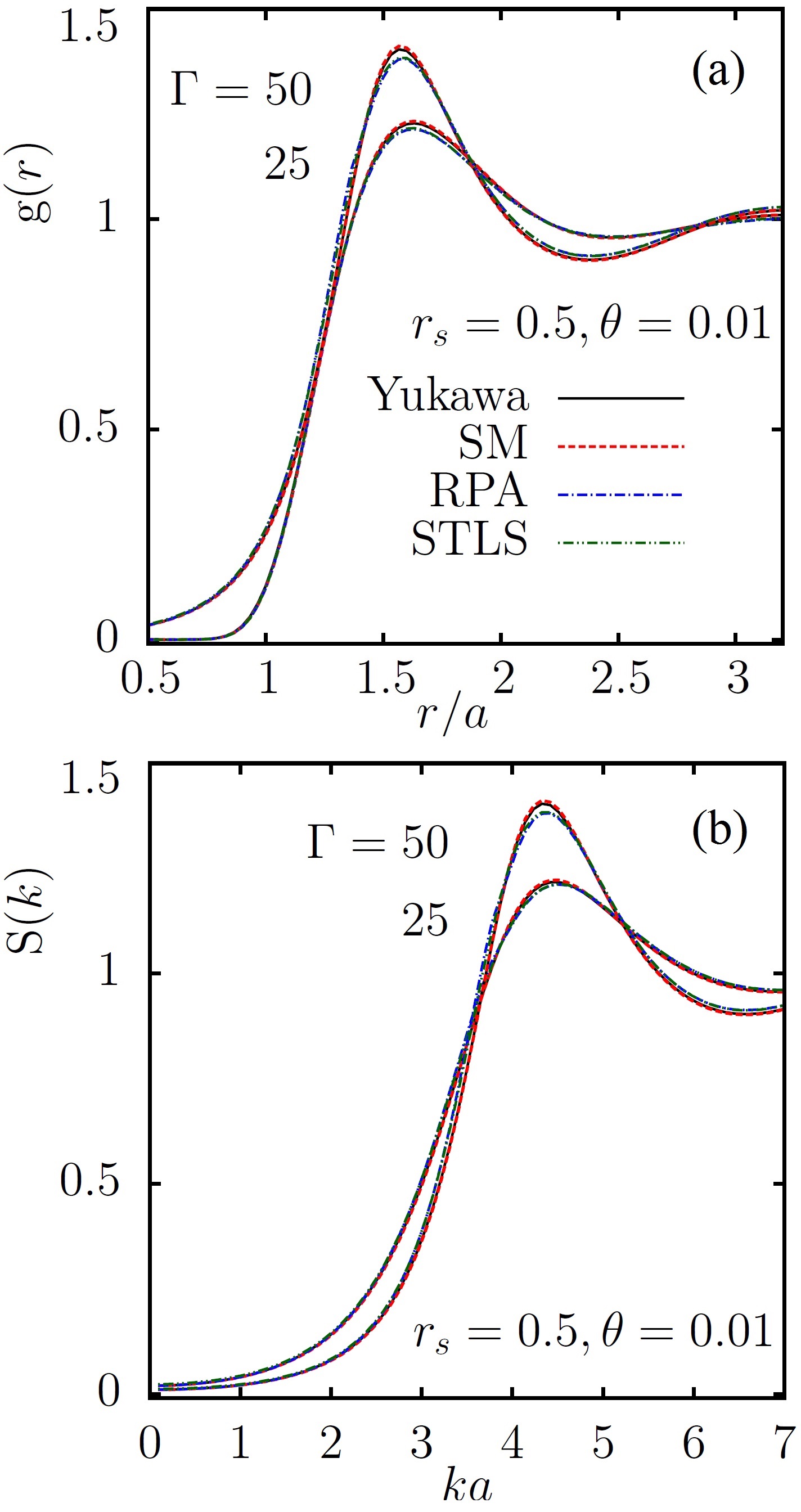}
\caption{Radial pair distribution function and structure factor computed on the basis of different screened potentials for different values of $\Gamma$ at $r_s=0.5$ and $\theta=0.01$.}
\label{fig:rs05theta0}
\end{figure}

\section{Structural properties of ions}\label{s:structure}

We now present results of calculations for three cases with different electronic correlations: $r_s\leq 1$, $r_s=2$, and $1<r_s<2$. 
These three cases can be characterized  as weakly, strongly, and moderately non-ideal regimes, respectively.
\subsection{Densities $r_s\leq 1$: weakly non-ideal electrons}
The case of the density parameter $r_s=1$ is especially interesting as it characterizes the transition from the regime where the RPA approach is justified to the regime where electronic non-ideality is important, and the RPA fails.
For a plasma with degenerate electrons, $\theta\ll1$, it defines the maximal density up to which the Thomas-Fermi potential approximated by Eq.~(\ref{Yukawa}) can be used. In Fig.~\ref{fig:potrs1}, 
 a comparison between different potentials is provided for the case of strong degeneracy, $\theta=0.01$, with $r_s=1$. 
It is clearly  seen that the electronic correlations (taken into account in the STLS screened potential) lead to a stronger screening of the ion potential.

In Fig.~\ref{fig:rs1theta0v2}  the RPDF calculated using different potentials at $\Gamma=0.5,~1,$ and 5 is presented.
The RPA screened, Yukawa, and SM potentials give almost the same result up to $\Gamma=10$, while the STLS screened potential, due to the stronger screening, produces a slightly smaller correlation-hole in the RPDF.
Surprisingly enough, at higher coupling parameters, the data obtained using the RPA screened potential  is much closer to the STLS result rather than to the results  calculated on the basis of the Yukawa and SM potentials, as
 illustrated in Fig.~\ref{fig:rs1theta0G25-50v2} for $\Gamma=25$ and $50$.  
This appears to be because the structural properties of the strongly coupled ions are sensitive to the exact shape of the potential.
We clarify this point by presenting the ratio of the considered potentials 
  to the Yukawa potential in Fig.~\ref{fig:mod_pot}. One can see that the RPA potential has oscillations around the decaying SM potential. 
  We note that, by design, the SM potential is a better approximation to the RPA than the Yukawa potential, but gives almost the same RPDF as the Yukawa potential.
  The STLS potential also has such oscillations, but with stronger overall screening. 
  In fact, these oscillations in the RPA and STLS potentials have the same shape (position of local extrema etc.) up to $r/a=3.75$.
  This can be seen by looking at the difference between the RPA screened potential and other potentials presented in Fig.~\ref{fig:dif_pot2}. 
  The difference in the shape of the oscillations in the STLS and RPA potentials appears only when the oscillations  turn 
  into the well-known Freidel oscillations with asymptotics $\sim \cos(2k_Fr)/r^3$ at $r/a>5$. Now, to show that the mentioned similarity in the pattern of the STLS and RPA potentials is the reason for the 
  agreement in the RPDF calculated on their basis, we introduce the following potential, which is a hybrid of the RPA and Yukawa potentials:
 \begin{equation}\label{hybrid}
  \Phi_i(r) =
  \begin{cases}
    \Phi_{\rm RPA} (r), & \text{ $r<r_i $} \\[2ex]
    \frac{\beta_i(r_i)}{r}\exp(-k_Y r), & \text{ $r\geq r_i$}
  \end{cases}
\end{equation}
 where $\beta_i(r_i)=\Phi_{\rm RPA} (r_i) r_i \exp(k_Y r_i)$, $r_1/a=5.5$, $r_2/a=2$, and $r_3/a=1.25$. 
 The values of $r_i$ are chosen such that at $r>r_1$ the oscillations of the RPA potential around zero (Fridel oscillations) take place, at $r>r_2$ the second maximum of the RPDF is located, and 
 at $r>r_3$ the first peak of the RPDF appears. Therefore, potential $\Phi_i$ (\ref{hybrid}) is the RPA screened potential up to $r_i$ and has the Yukawa-type screening part, for $r_i>r$. Using this potential allows us
 to eliminate the effect of the oscillations from different regions one by one and, thereby, to check how sensitive the structural properties of the strongly coupled ions are  to these oscillations. 

In Fig.~\ref{fig:gr_osc50}, the RPDF calculated using  $\Phi_i$ (denoted as $\rm {RPA}_i$) is compared to the RPDF of the YOCP. 
The potential $\Phi_1$ gives the same result as the RPA potential, the potential $\Phi_2$ gives a result closer to the YOCP at $r/a>2$ and in agreement with the RPA  potential based result at $r/a<2$, and 
 the potential $\Phi_3$ closely reproduces calculations using the Yukawa potential (with small differences due to  the matching with the RPA potential at $r=r_3$). 
 Therefore, it is clear that the difference, at large coupling parameters, 
 in the RPDF computed using the RPA screened potential and the Yukawa (or SM) potential appears due to the manifestation of the positive oscillations of the RPA potential. 
 
   \begin{figure}[h]
 \includegraphics[width=0.48\textwidth]{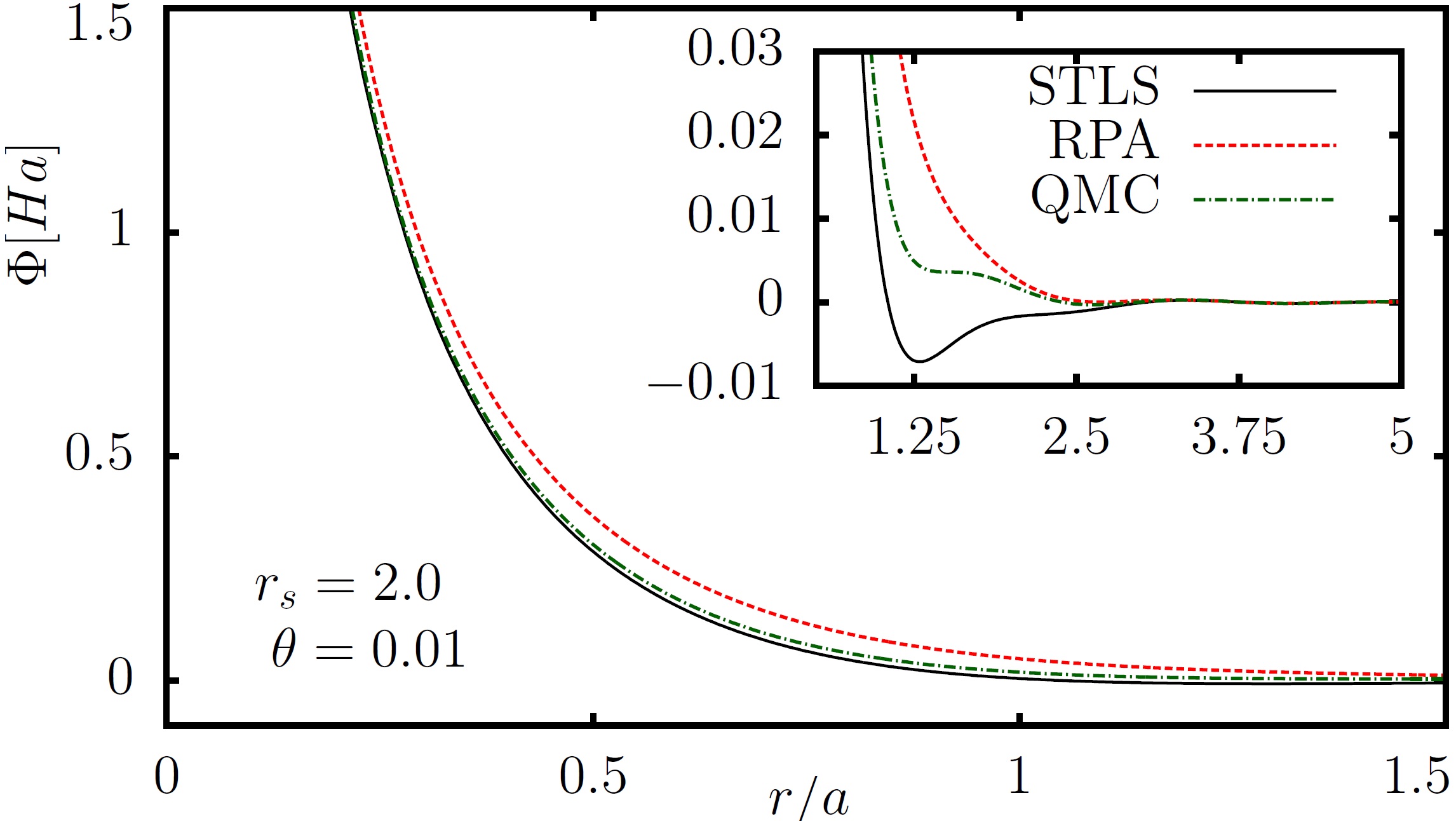}
 \caption{The STLS and RPA screened potentials in comparison with the QMC data based potential at $r_s=2.0$ and $\theta=0.01$.}
 \label{fig:potrs2}
 \end{figure}
 
 Next, we consider the static structure factor $S(k)$ (SSF). In Fig.~\ref{fig:rs1Sk}, the values of $S(k)$ are shown for $\Gamma$ in the range from 0.5 to 50. 
 Due to stronger screening, the STLS based results have a higher value of $S(0)$ in comparison with the case when the electronic correlations are neglected. This means that the inclusion of the effect 
 of non-ideality of the electrons results in a larger  isothermal compressibility of the ions.  
 At $ka< 3$, the RPA result is in very good agreement with the results obtained using the Yukawa and SM potentials for all considered coupling parameters. This means that the Fridel oscillations 
 do not affect the structural properties of the ions. This remains correct for all  $r_s\leq 2$ regardless the value of $\theta$.  
  Additionally, at $ka>2.5$, the STLS result is in good agreement with the results obtained using the RPA, Yukawa, and SM screened potentials up to $\Gamma=10$.
  At $\Gamma>10$ and $ka>3$, the RPA result remains in agreement with the STLS result, but not with the  Yukawa and SM screened potentials based results.
  This, again, is due to the aforementioned effect of the similarity in the oscillations of the RPA and STLS screened potentials. The latter is known to be associated with
  the so-called Kohn anomaly \cite{Kohn, Else}, i.e., a local non-monotonicity of the dielectric function around $k\sim 2 k_F$. 
  
  After the detailed consideration of the case of strongly degenerate electrons, we consider the impact of the thermal excitations of electrons at $r_s=1$.
  In Fig.~\ref{fig:gr_theta}, the values of $g(r)$ and $S(k)$ are shown at different degeneracy parameters, $\theta=0.1,~0.2$, and 0.3.
  From this figure it is clear that the thermal effect leads to the suppression of all electronic quantum non-locality and non- ideality effects at $\theta>0.3$.
  Therefore, in the case of $r_s=1$,  at $\theta>0.3$ the simple Yukawa potential (\ref{Yukawa}) provides a fairly good description of the structural properties of the strongly coupled ions.
  
  With increase in the density, the role of the correlations of the electrons and of the quantum oscillations related to the Kohn anomaly diminishes.
  This is confirmed by the calculations of $g(r)$ and $S(k)$ presented in Fig.~\ref{fig:rs05theta0} for $r_s=0.5$ and $\theta=0.01$. We see that 
  the results obtained using different potentials are in good agreement with each other up to $\Gamma=25$.   
  The effect of the oscillations in the STLS and RPA screened potentials leads to a slight difference around the first peak at $\Gamma=50$.  
 Additionally, it should be noted that the considered effect of the oscillations remains valid at $r_s\leq1$, 
 but the conclusion that with increase in the coupling parameter ions become more sensitive to the features of the 
 screened potential is general.
 
   \begin{figure}
 \includegraphics[width=0.43\textwidth]{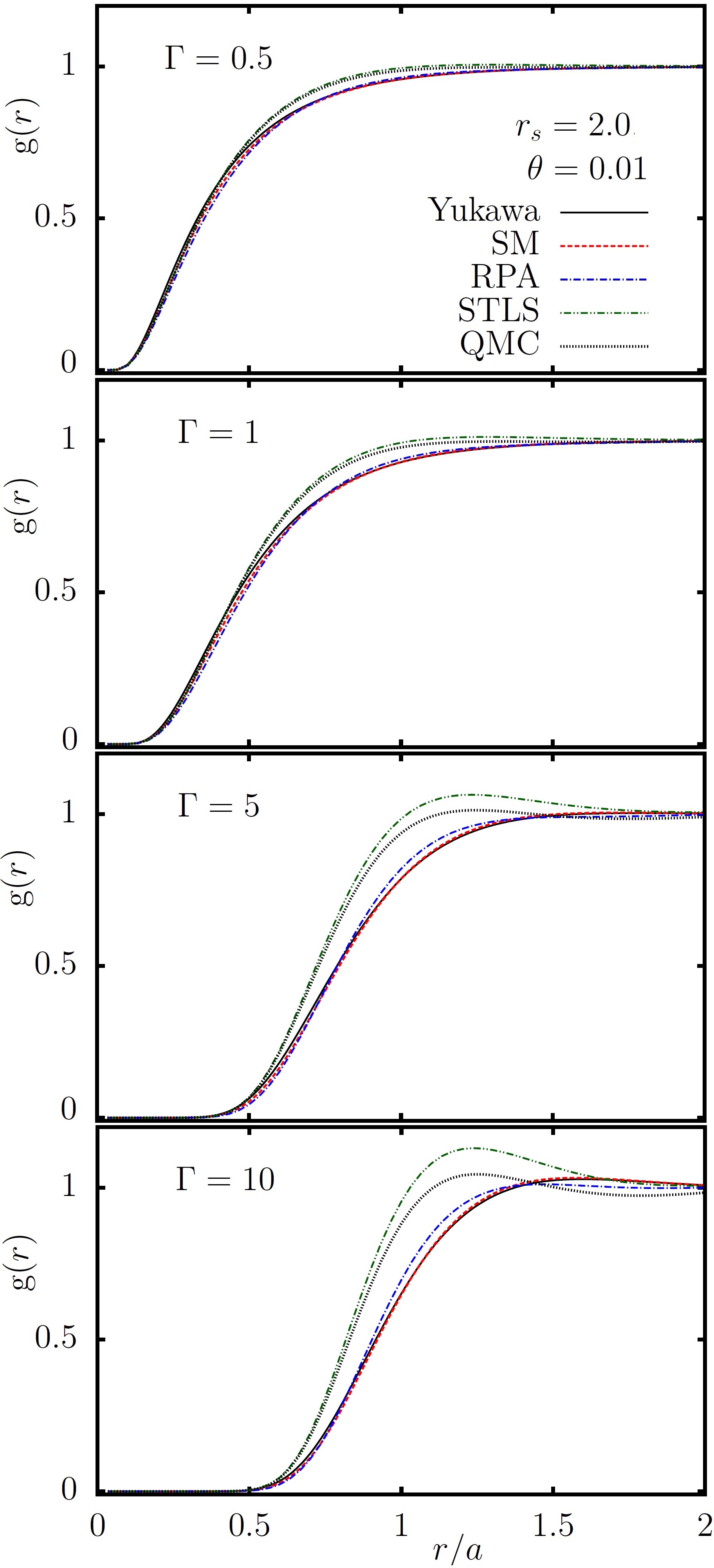}
 \caption{Radial pair distribution function of ions at coupling parameters in the range from 0.5 to 10 computed using different approximations for the screened ion potential.}
 \label{fig:rs2theta0}
 \end{figure}

    \begin{figure}
 \includegraphics[width=0.45\textwidth]{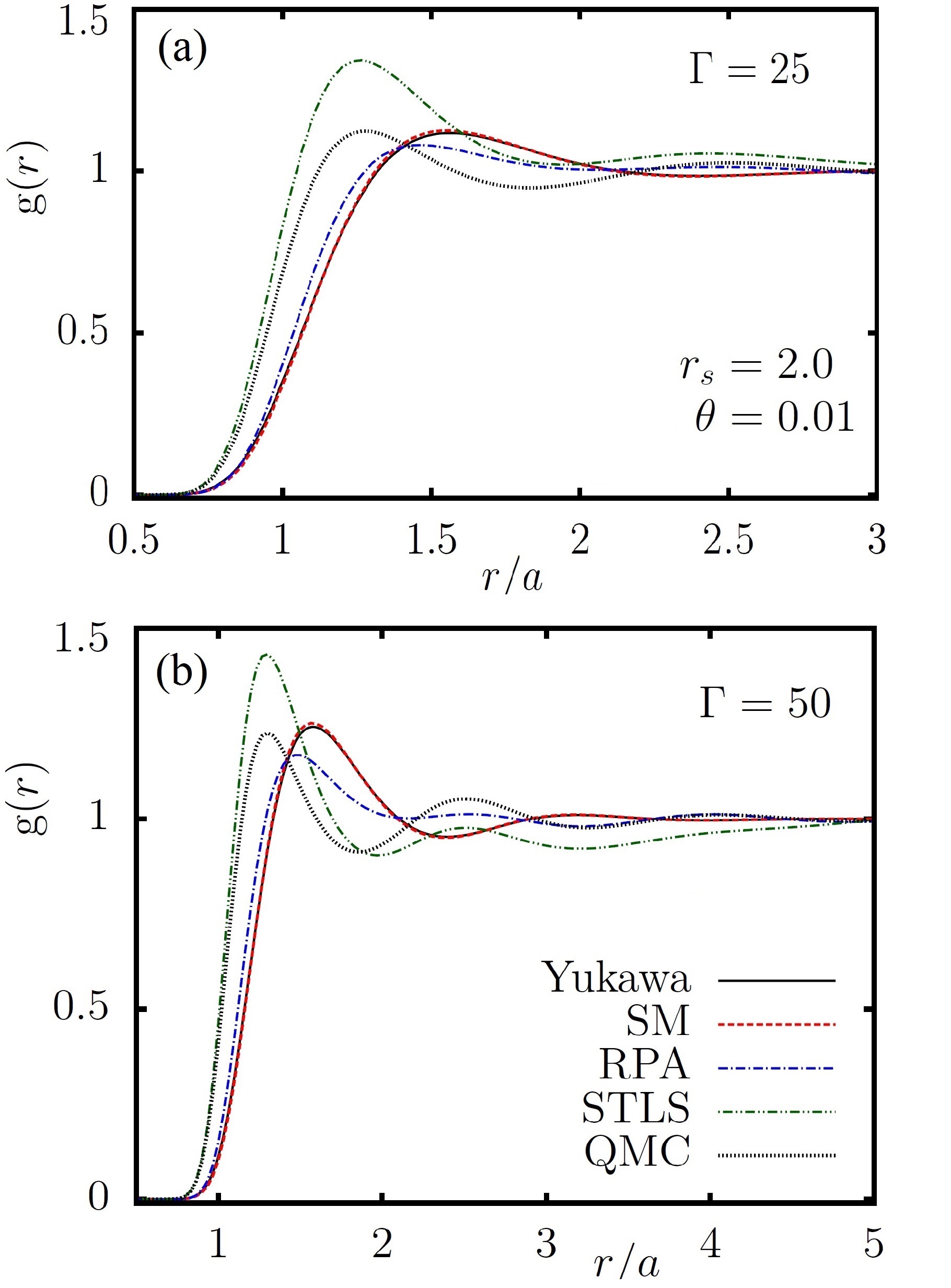}
 \caption{The same as in Fig.~\ref{fig:rs2theta0} but for $\Gamma=25$ and 50.}
 \label{fig:rs2theta0G25v2}
 \end{figure}
 
      \begin{figure}
 \includegraphics[width=0.45\textwidth]{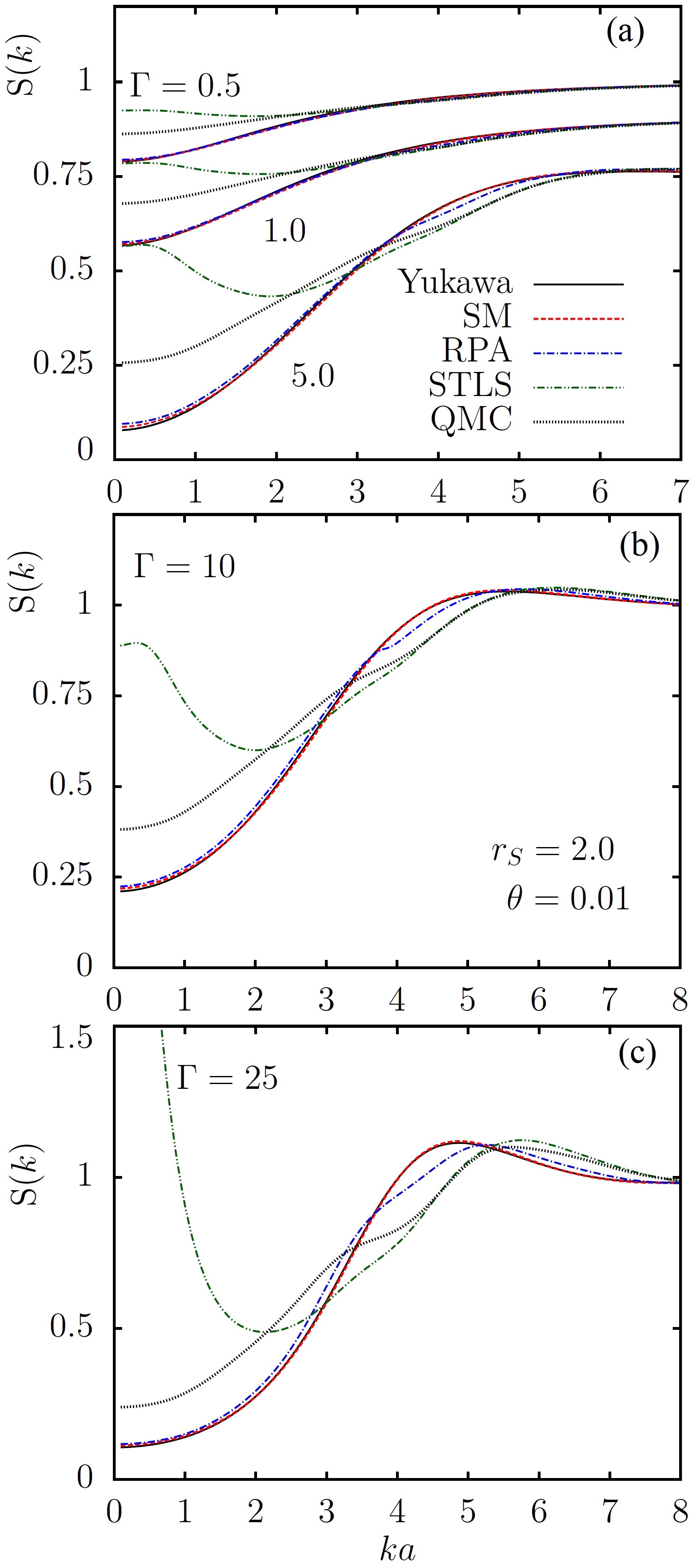}
 \caption{Static structure factor computed using various screened ion potentials at different coupling parameters.}
 \label{fig:rs2Sk}
 \end{figure}

     \begin{figure}
 \includegraphics[width=0.48\textwidth]{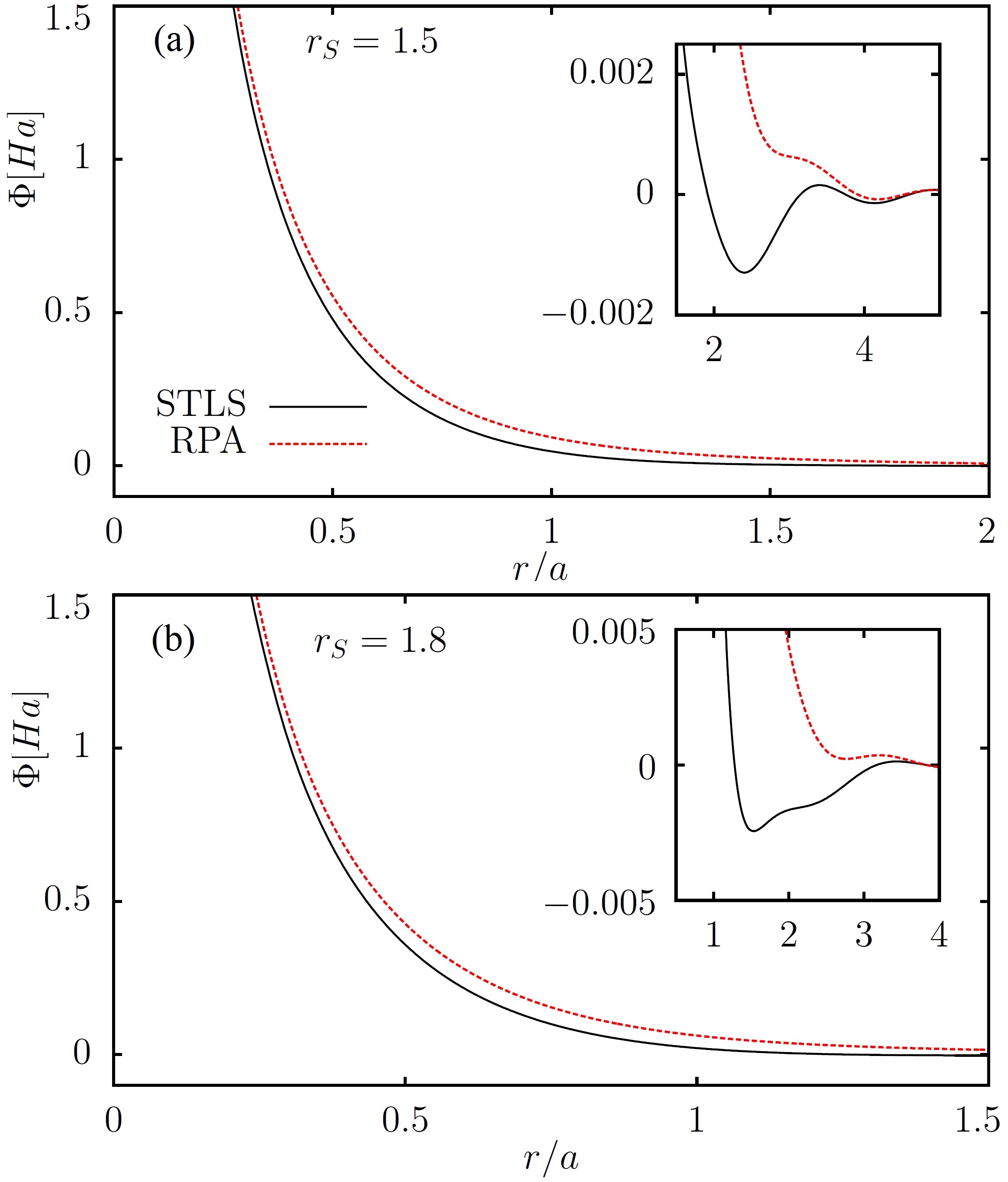}
 \caption{The STLS and RPA screened potentials at $r_s=1.5$ and $r_s=1.8$ ($\theta=0.01$).}
 \label{fig:potrs15rs18}
 \end{figure}

 \subsection{Density $r_s=2$: non-ideal electrons}

  The effect of the electronic non-ideality is expected to be more important as $r_s$ increases. For the case $r_s=2$, the STLS results can be checked against more reliable ground state QMC data \cite{Corradini}. 
  
  As it was mentioned above, we use the accurate parametrization by Corradini \textit{et al.} \cite{Corradini}. 
  It is important to note that in many  previous and recent works (e.g. \cite{Reinholz95, SjostromPRB}),  parametrizations like those by Ichimaru and Utsumi \cite{Ichimaru} (which is based on STLS),
  were used with the restriction $G(k\to \infty)<1$ that originated from the earlier works of Shaw \cite{Shaw} and Niklasson \cite{Niklasson}.
  However, as was first  shown by Holas \cite{Holas}, it turned out that the condition $G(k\to \infty)<1$ is invalid and the correct asymptotic behavior is $G(k\to \infty)\sim k^2$.
  This was confirmed by the ground state QMC simulation results in Ref.~\cite{Moroni}. Recently, 
 first \textit{ab initio} calculations of the static density response function of electrons at finite temperature have been successfully
performed \cite{dornheim17, groth17}, but the $k$-resolution 
is not yet sufficient  for the implementation into the calculation of the screened ion potential.   

 In Fig.~\ref{fig:potrs2}, the STLS, RPA, and QMC-based screened ion potentials are presented for $r_s=2$ and $\theta=0.01$.
 At $r/a\leq 1$, the STLS potential is in good agreement with the QMC data based potential. At $r/a\simeq 1.25$, in contrast to the RPA and QMC data based potentials, the STLS screened potential has
  a negative minimum.  In Fig.~\ref{fig:rs2theta0}, the corresponding RPDFs,  $g(r)$, are shown for $\Gamma$ in the range from $0.5$ to $10$. At $\Gamma\leq1$, the STLS result is in agreement with the QMC data based result. 
  As the ionic coupling parameter increases, at $\Gamma\geq 5$, the negative minimum in the STLS potential  leads to a deviation of $g(r)$ calculated using the STLS potential from $g(r)$ obtained using 
  the QMC data based potential.  This disagreement is more pronounced at $\Gamma=25$ and 50 as it is shown in Fig.~\ref{fig:rs2theta0G25v2}, where the ion-ion attraction due to the negative minimum 
  in the STLS potential leads to a much higher first peak in $g(r)$.  As it is expected, the RPA screened, Yukawa, and SM potentials  
  neglecting non-ideality of electrons are not in agreement with results obtained using the QMC data based potential. Additionally, 
   the Yukawa and SM potentials appear to be poor approximations to the RPA screened potential for the description of the RPDF  at $\Gamma>5$.
   
   In Fig.~\ref{fig:rs2Sk}, the static structure factor calculated using different potentials at $\Gamma=0.5,~1.0,~5.0,~10$ and 25 is shown.
   In this figure, the emergence of a minimum at $ka\simeq 2$  with increase in $\Gamma$ is demonstrated. 
   This minimum is the result of the attractive part in the STLS screened ion -ion interaction potential (see Fig.~\ref{fig:potrs2}). 
   As it is illustrated  in Fig.~\ref{fig:rs2Sk}, it is crucial that this feature of $S(k)$ is not confirmed by the calculations based on the more accurate QMC based ion-ion interaction potential. 
   Therefore, modeling screening in the STLS approximation essentially fails to provide a correct description of $S(k)$ and $g(r)$ at these parameters, showing unphysical effects of attraction between ions. 
   However, from Fig.~\ref{fig:rs2Sk} it is also evident that  the STLS results and the QMC data based results at $ka>4$ (out of the region of the unphysical minimum) are in good agreement with each other. 
   This is because the STLS potential correctly reproduces the QMC data based potential at $r/a<1$ (see Fig.~\ref{fig:potrs2}).
   In this way, the detailed examination employing  the QMC static local field correction allowed us to find the reason for the failure of the STLS screened potential, at certain plasma parameters, in description of the structural properties of the strongly coupled ions
    in a dense quantum plasma. 
   
  \subsection{Density $1<r_s<2$: moderately non-ideal electrons}
 
 Considering densities corresponding to $1<r_s<2$ we can reveal more information about the applicability of the STLS screened potential for the calculation of $g(r)$ and $S(k)$, at $\Gamma>1$.
 The STLS and RPA potentials are shown in Fig.~\ref{fig:potrs15rs18}, for $r_s=1.5$ and $1.8$, at $\theta=0.01$. 
 Again, the STLS potential is more strongly screened than the RPA potential. Additionally, at $r/a>1$, the STLS  potential has a negative minimum at both densities, $r_s=1.5$ and $r_s=1.8$, but with different 
  overall behavior after the minimum. At $r_s=1.5$, the STLS potential has an oscillatory pattern with alternating positive and negative extrema and, at $r_s=1.8$, a well developed region of attraction is clearly seen.
  This difference leads to dramatic consequences in the structural properties \cite{Matsuda, Canales}. 
  
\begin{figure}
 \includegraphics[width=0.45\textwidth]{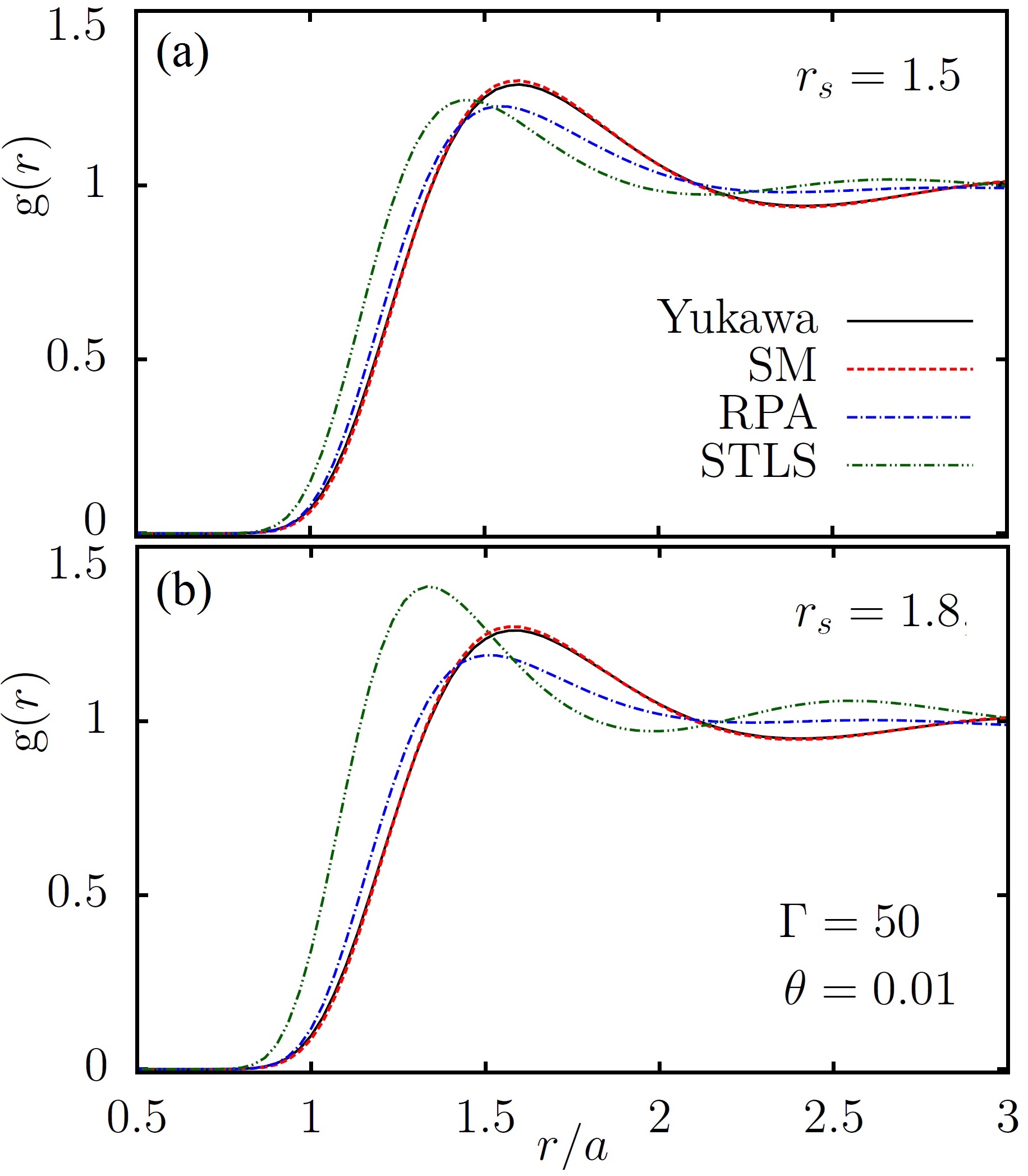}
 \caption{Radial pair distribution function of ions for $r_s=1.5$ and $1.8$ at $\theta=0.01$ and $\Gamma=50$.}
 \label{fig:gr_rs18rs15}
 \end{figure}

\begin{figure}
 \includegraphics[width=0.45\textwidth]{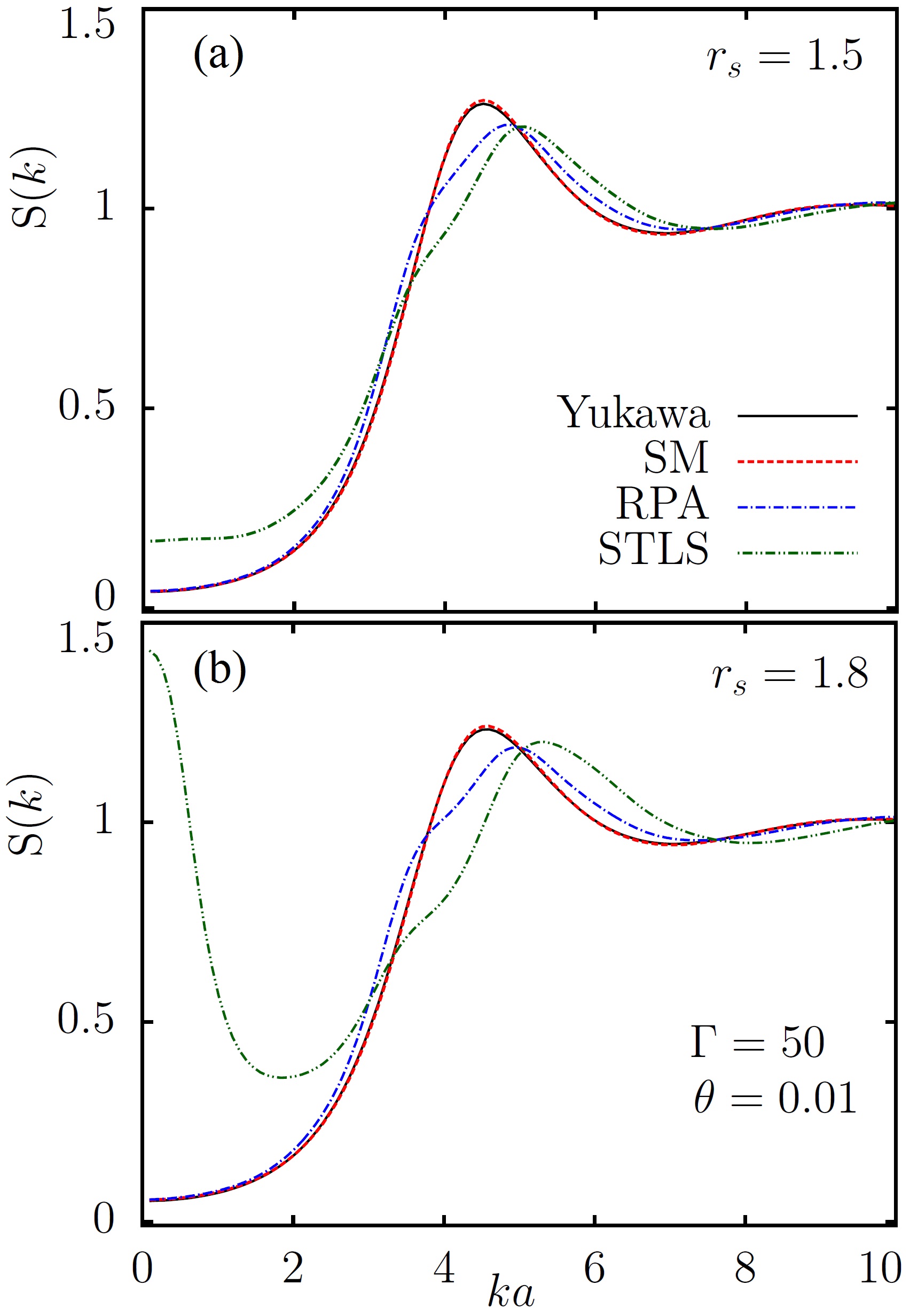}
 \caption{Structure factor of ions for $r_s=1.5$ and $1.8$ at $\theta=0.01$ and $\Gamma=50$.}
 \label{fig:sk_rs18rs15}
 \end{figure}

 In Fig.~\ref{fig:gr_rs18rs15}, the RPDF  of ions at $r_s=1.5$ and $r_s=1.8$ is shown for $\theta=0.01$  and $\Gamma=50$. 
At $r_s=1.5$, due to stronger screening, the RPDF calculated using the STLS potential has a smaller correlation-hole and a lower peak than the RPDF computed on the basis of the Yukawa potential.
In contrast, due to the attraction part in the screened potential, at $r_s=1.8$ the peak in the RPDF of ions interacting through the STLS screened potential 
is much  higher than that of the YOCP with pair potential (\ref{Yukawa}). Such an effect is not visible at $r_s=1.5$ because the asymptote of the potential has a pattern of oscillations around zero with closely 
enough situated extrema so that the effect of the repulsive and attractive parts of the pair potential mutually compensated \cite{Matsuda}, similar to the case $r_s=1$ (see Fig.~\ref{fig:dif_pot2}). 
In Fig.~\ref{fig:sk_rs18rs15}, corresponding values of $S(k)$  are shown. 
The manifestation of the unphysical behavior at $r_s=1.8$ 
due to ``uncompensated'' effect of the attraction between ions interacting via the STLS screened potential is more clear in $S(k)$ at $ka<4$, while, at $r_s=1.5$, such a feature of $S(k)$ is absent. 
Additionally, we note that at $\Gamma=50$, $\theta=0.01$, $r_s=1.5$, and $r_s=1.8$ the Yukawa and SM potentials fail to correctly reproduce the results found using the RPA screened potential.

\begin{figure}
 \includegraphics[width=0.45\textwidth]{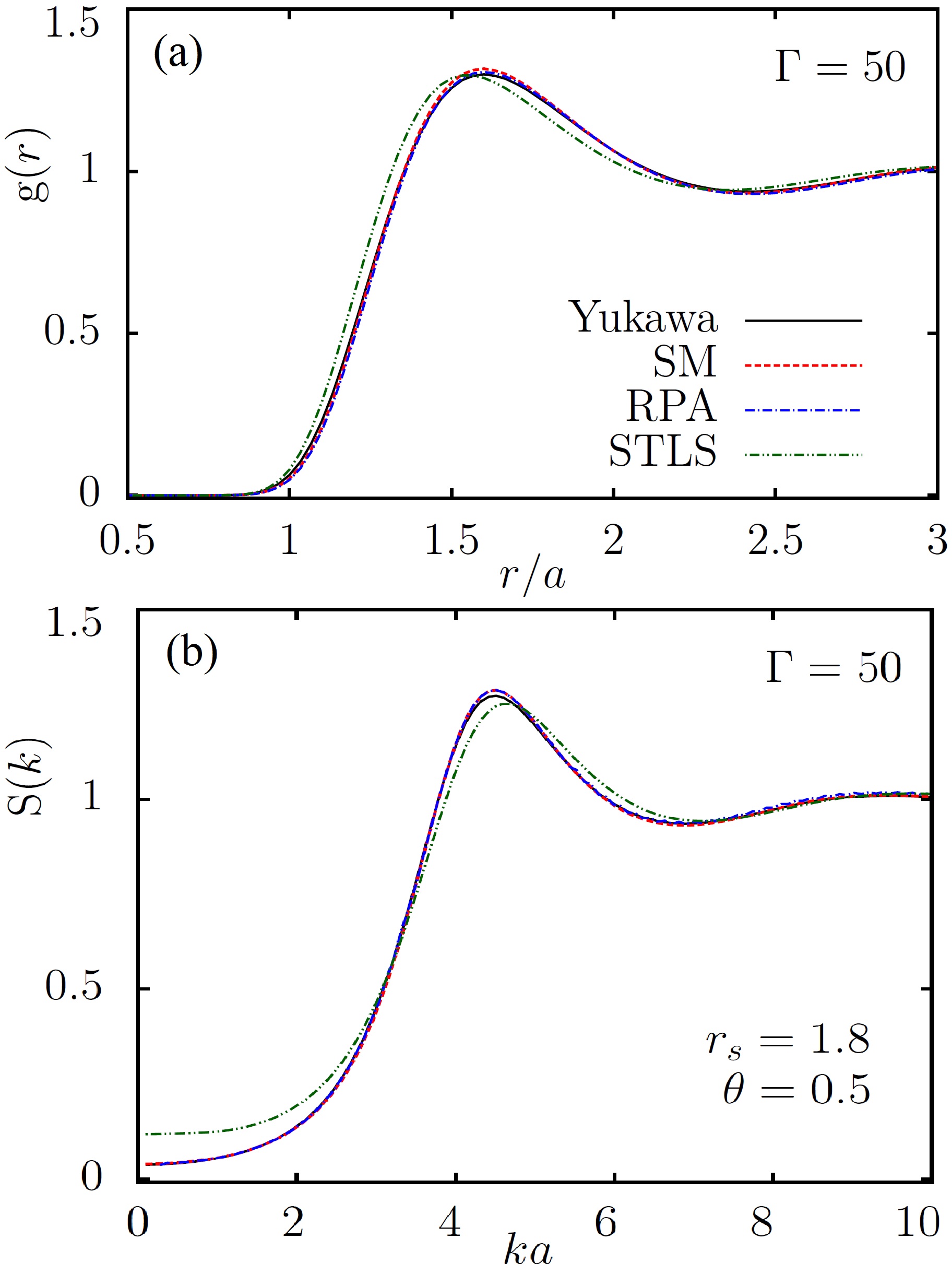}
 \caption{Radial pair distribution function and structure factor of ions at $\theta=0.5$, $r_s=1.8$, and $\Gamma=50$.}
 \label{fig:g_S_theta05rs18}
 \end{figure}

As it is expected, the demonstrated feature of $S(k)$ appearing as the result of the attraction part in the STLS potential at $r_s=1.8$ is suppressed at larger $\theta$ due to thermal electronic excitations.
This is illustrated in Fig.~\ref{fig:g_S_theta05rs18}, where $g(r)$ and $S(k)$ are shown at $r_s=1.8$, $\theta=0.5$, and $\Gamma=50$. In Fig.~\ref{fig:g_S_theta05rs18} the effect of the attraction is not visible at all.  
Further,  at $\theta=0.5$, we find good agreement between the results obtained on the basis of the RPA, Yukawa, and SM potentials (being different from the STLS based result), 
meaning the higher order quantum effects, which are neglected in the Yukawa and SM potentials, are diminished as well.

     \begin{figure}
 \includegraphics[width=0.43\textwidth]{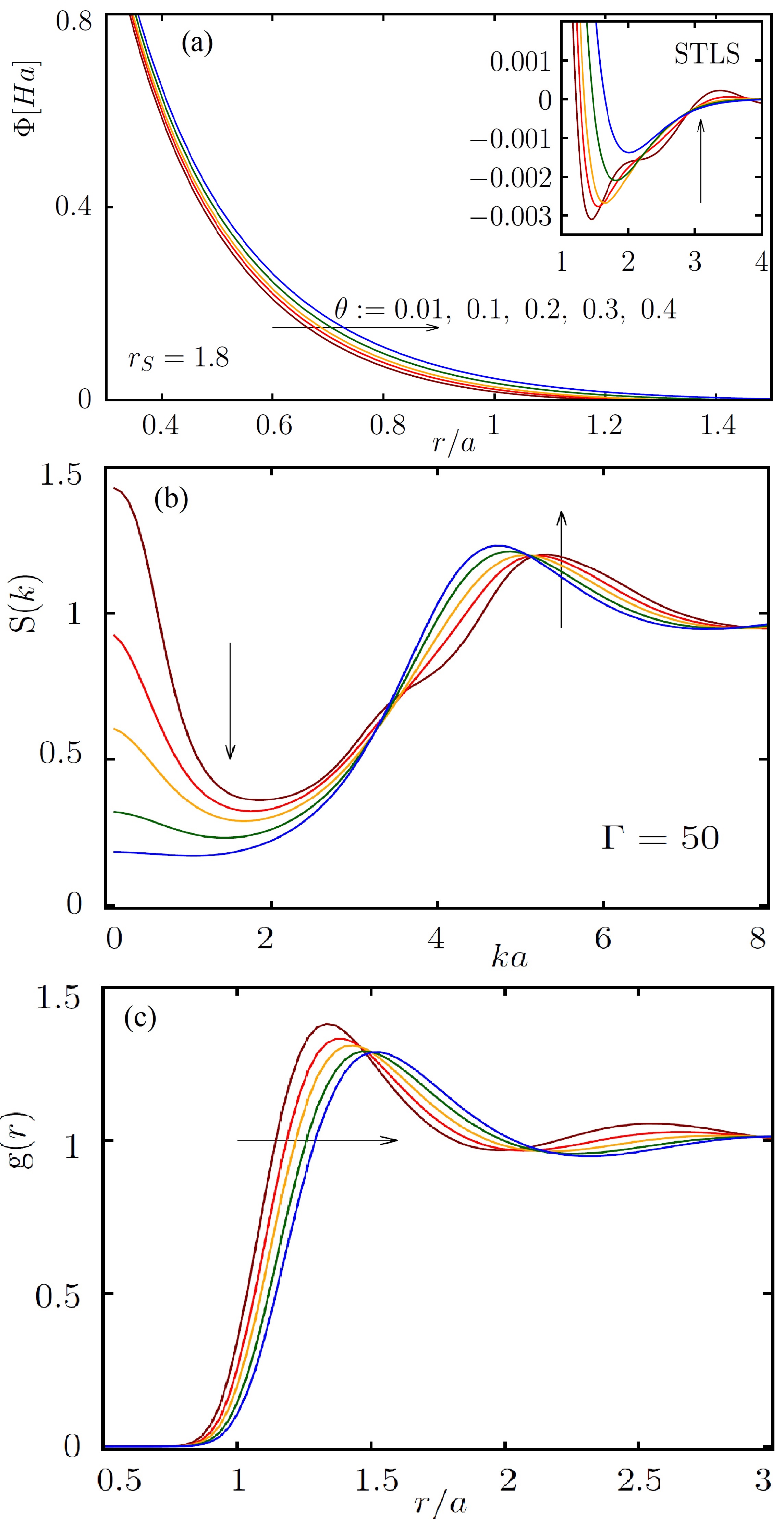}
 \caption{The screened ion potential (a), static structure factor (b), and radial pair distribution function (c) at $\Gamma=50$ calculated using the STLS screened potential at $r_s=1.8$ and different values of the degeneracy parameter of electrons $\theta$.}
 \label{fig:potrs18theta}
 \end{figure}
 
   \begin{figure}[h]
   \hspace*{0.5 cm}
 \includegraphics[width=0.4\textwidth]{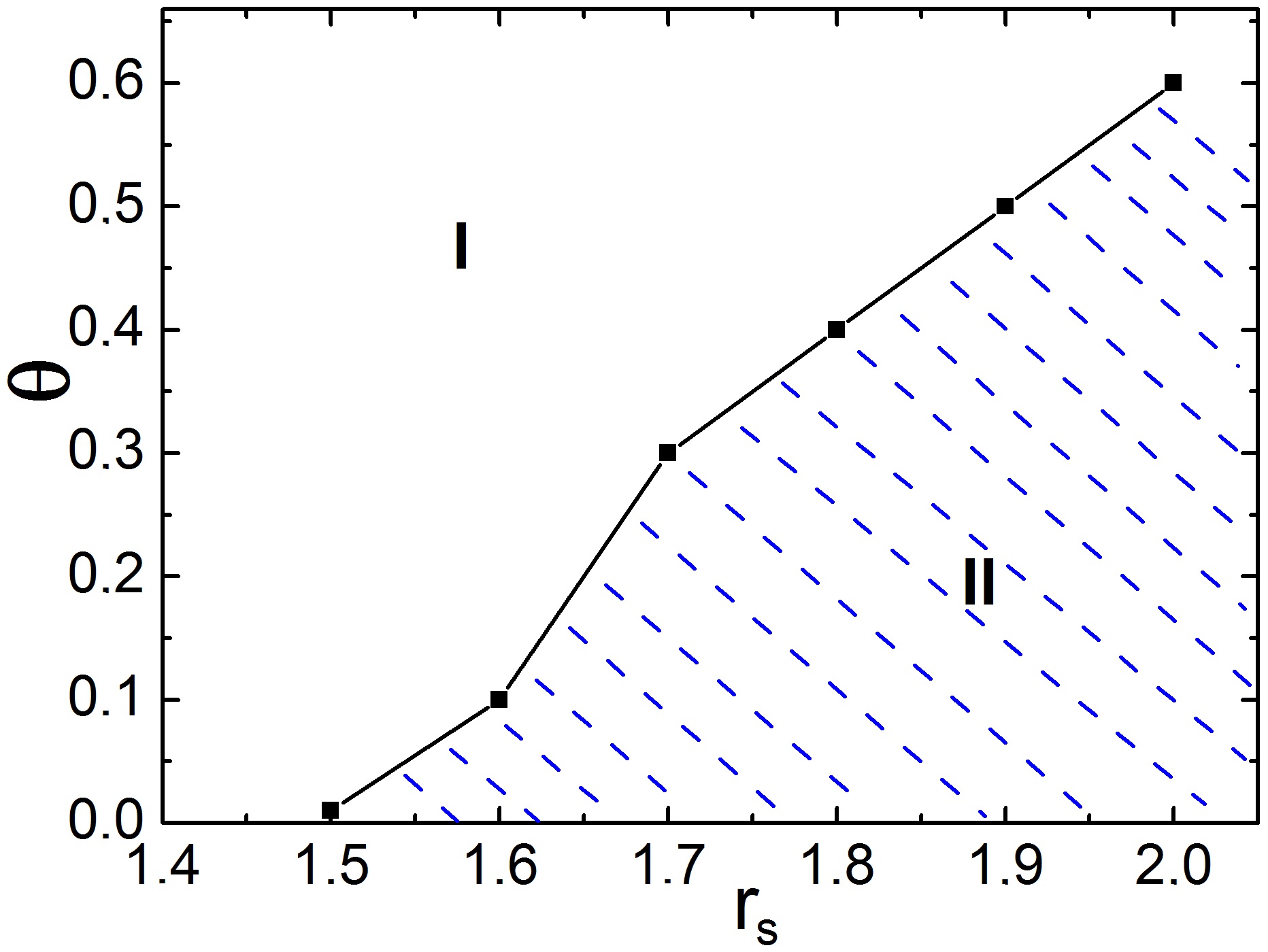}
 \caption{In the region $\rm I$ the artifact feature of $S(k)$ due to the attractive part in the STLS screened potential does not appear, and 
 in the region ${\rm II}$ (dashed area) the unphysical absolute minimum in $S(k)$ at $k>0$  builds up at $\Gamma\geq 1$.}
 \label{fig:region_applica}
 \end{figure}
 
  We verified the appearance of the feature in $S(k)$ and $g(r)$, which is related to the attractive part of the STLS potential, by performing independent MD simulations. The MD simulations 
  confirmed our findings from the solution of the  Ornstein-Zernike equation with the HNC closure, see Appendix B. 
 
 \subsection{Applicability range of the STLS screened ion potential}

 We can now use the absence of the pronounced impact of the attraction in the ion-ion interaction on $S(k)$ as  a necessary criterion for the applicability of the STLS screened potential for the description 
 of the structural properties of the dense plasma with strongly coupled ions. To proceed further, we consider different values of the degeneracy parameter. 
 In the top panel of Fig.~\ref{fig:potrs18theta}, the STLS potential is shown at $\theta=0.01,~0.1,~0.2,~0.3$ and 0.4. 
 With increasing degeneracy parameter, the screening becomes weaker, the absolute values of the negative minimum 
  decrease, and the position of the negative minimum shifts to a larger distance. 
 Corresponding $S(k)$ and $g(r)$ of the ions at $\Gamma=50$ are presented in the middle and the bottom panels of Fig.~\ref{fig:potrs18theta}, respectively.
 The electronic thermal effect results in the suppression of the manifestation of the ion-ion attraction in $S(k)$ at $ka<4$. 
  At $\theta=0.4$, the value of $S(k)$ at the minimum (located in the range $0<ka<2$) differs by less than $5\%$ from $S(0)$.
  In fact, this difference rapidly disappears with increasing $\theta$. Indeed, the minimum in $S(k)$ at $ka<4$ does not exist already at $\theta=0.5$ (see Fig.~\ref{fig:g_S_theta05rs18}). 
  From Fig.~\ref{fig:potrs18theta} one can see that the height of the first peak decreases with increase in $\theta$ due to the weakening of the attractive part of the potential, while the correlation-hole increases due to weaker screening.
 
 To determine parameters at which the STLS potential can be used for the description of strongly coupled ions, we realized a large scale study of $S(k)$ 
 at $\Gamma\leq100$, $0.01\leq \theta\leq1$, $r_s\leq 2$. The results are summarized in Fig.~\ref{fig:region_applica}, where two regions in the $\theta-r_s$ plane are indicated. In region $\rm I$,
   the artificial feature of $S(k)$ due to the attractive part in the STLS potential does not appear. In contrast, in region ${\rm II}$ the unphysical absolute minimum in $S(k)$ at $k>0$
   builds up at $\Gamma\geq 1$.  In region $\rm I$ of Fig.~\ref{fig:region_applica}, at $r_s\leq1.5$, the STLS potential has oscillatory asymptote and closely enough located repulsion and attraction parts exhibit a mutually compensating effect 
  leading to monotonic decay of $S(k)$ as $k$ decreases. For instance, the same is true regarding the oscillations around zero in the RPA potential (Fridel oscillations). In contrast, in region $\rm I$, at $r_s>1.5$ the role of the attractive part of 
   the STLS potential diminishes due to electronic thermal effects.

 \section{Summary and Outlook} \label{s:dis}

 From our analysis of the structural properties of strongly coupled ions on the basis of different screened ion potentials---at typical parameters of non-ideal quantum plasmas---we have the following conclusions:\\
 
 1.~  We determined the region of densities and temperatures where the STLS  description of screening by partially (or totally) degenerate 
 electrons can be used for the calculation of the structural properties of ions. 
 At $r_s>1$,  electronic correlations beyond RPA have a non-negligible effect on the structural properties of strongly coupled ions in quantum plasmas.
 In particular, correlations (non-ideality) of the electrons result in a larger  isothermal compressibility of the ions due to the stronger screening of the ion charge.\\
 
 2.  The applicability of the simplest Yukawa model was gauged by comparing with the results obtained using a more general description of the screening by the quantum  random phase approximation. 
 It was shown that at $\theta\simeq 0.1$  the Yukawa and  SM potentials can not be used as a reliable approximation to the RPA potential at $\Gamma>10$ and $r_s>1$. 
 This was explained by the use of the  long wavelength limit of the density response function in the entire $k$ range in the case of the Yukawa and SM potentials.  
 At these plasma parameters, such an approximation appears to be justified when the manifestation of the Kohn anomaly is suppressed by the thermal excitations of electrons. It is interesting that at $\Gamma\leq50$, $r_s\leq2$ and $\theta\geq0.01$ 
 the somewhat more complex (but still analytical)  SM potential gives the same result as the simpler Yukawa potential when applied for the calculation of the radial pair distribution function and the structure factor of the ions. \\
  
  3. Calculations using different screened potentials clearly show that strongly coupled ions at $\Gamma\geq10$ can be very sensitive to the peculiarities of the shape of the pair interaction potential and, therefore, 
  to the approximation used for the description of screening by electrons.\\
   
   4. It was revealed that the HNC approximation can be used for an accurate description of the structural properties of the strongly coupled particles 
   up to  a maximal coupling parameter the value of which depends on the screening strength.
   In the example of the Yukawa potential, the maximal effective coupling  is $\Gamma_{\rm eff}\simeq 10$. 
   This is a very useful finding as the HNC approximation is often used as a fast and easy way of 
   incorporating ionic non-ideality effects into various theoretical descriptions. \\

\begin{figure}[h]
   \hspace*{0.5 cm}
 \includegraphics[width=0.45\textwidth]{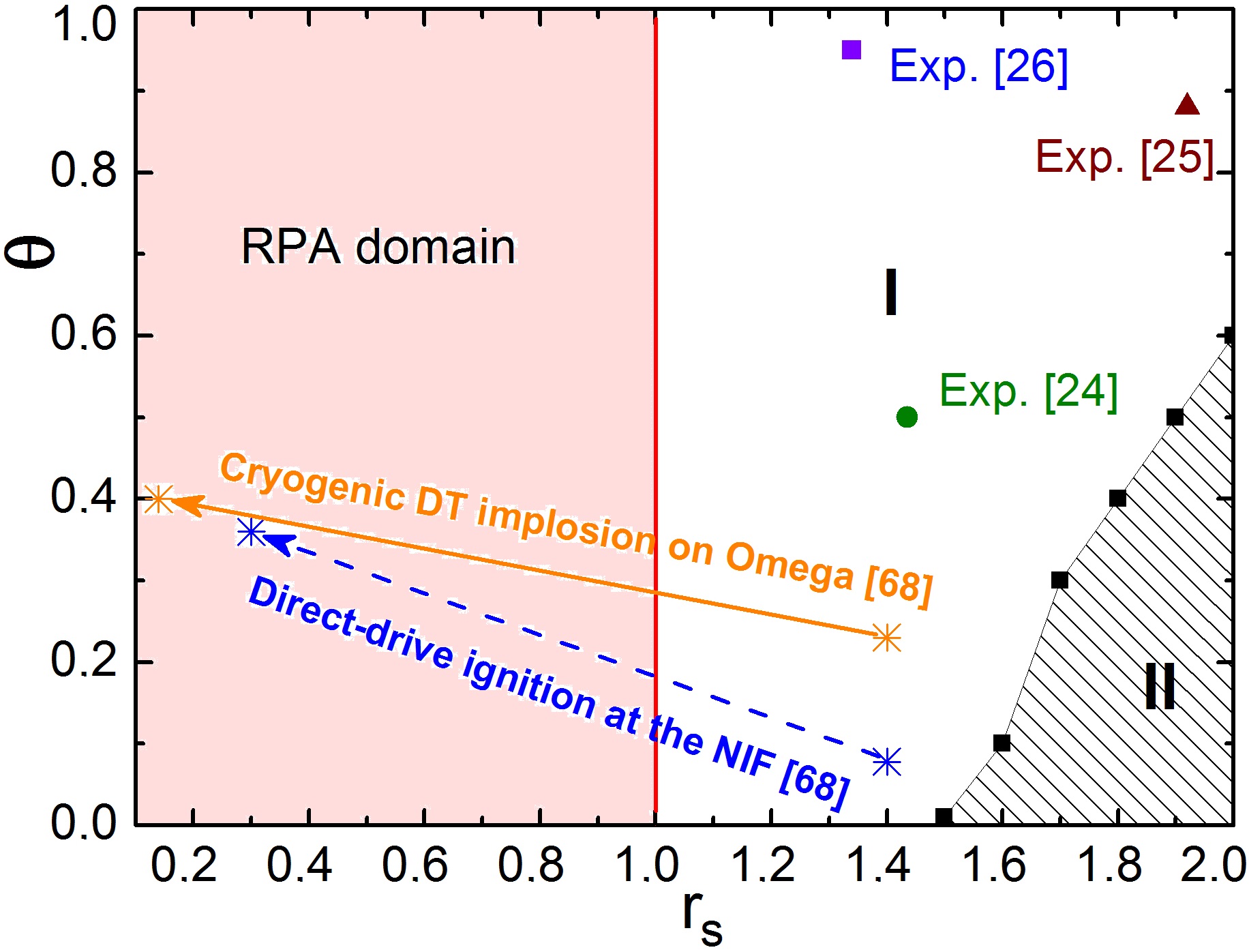}
 \caption{The same as in Fig. 22, but with  examples of experiments where quantum plasmas with strongly coupled ions were realized. Additionally, the RPA domain is indicated. 
 The path which the plasma with non-ideal ions undergoes in ICF experiments is given approximately  on  the  basis 	of the data	extracted from  Ref.~\cite{Strong_Coupling_and_Degeneracy_in_ICF}.}
 \label{fig:exp}
 \end{figure}

 Let us briefly discuss the possible application of the results of the present work. In Fig.~\ref{fig:exp}, some of the experimentally obtained plasmas with parameters overlapping with those considered  in  this  work  ($\theta<1$, $r_s<2$, and $\Gamma>1$) are shown  on  the $\theta-r_s$ plane. Further, a part of the paths which the plasma with $\Gamma>1$ undergoes in ICF experiments at the NIF and Omega  are  sketched, based on data from Ref.~\cite{Strong_Coupling_and_Degeneracy_in_ICF}.
  In addition we have marked the domains of applicability of RPA and STLS, respectively, as discussed in this paper. Note that the border of the RPA domain  should be understood qualitatively. The strict applicability condition in a fully degenerate plasma is $r_s\ll 1$. 
  At the considered temperatures with $\theta<1$, this condition is softened by thermal excitations, allowing one to use the RPA closer to $r_s=1$.
       Figure~\ref{fig:exp} shows that the STLS approach discussed in Sec.~\ref{s:theory} 
    is applicable to ICF plasmas starting from the initial stage, where electronic correlations and quantum non-locality effects  are  crucial ($r_s>1$ and $\theta \sim 0.1$), 
    up  to  the  final regime where electrons are ideal ($r_s\ll1$). 

  An  accurate computation of the electrodynamic and transport properties of dense plasma is important, e.g., for the understanding of the processes occurring during fuel compression in ICF experiments.  
   For the investigation of the plasma at the discussed parameters, the ionic static structure factor is needed for the accurate calculation of the dynamic electron-ion collision frequency, 
   which can then be used to study the electrodynamic (optical)  properties of the plasma  \cite{PhysRevE.94.013203} (e.g., reflection and absorption coefficients, emission, and plasma stopping power). 
   Additionally, the ionic static structure factor is needed for the calculation of the plasma resistivity on the basis of the Rousseau-Ziman formula \cite{Bennadji}.
   The STLS screened potential can be used in MD simulations for the computation of the ionic dynamical structure factor and transport properties such as ionic viscosity and diffusion coefficient.
The ionic dynamical structure factor is needed for the description of the elastic scattering of x-rays off electrons in x-ray Thomson scattering experiments  \cite{Exp}.
   Note that  the  RPDF obtained using the HNC on the basis of the STLS potential allows one to calculate plasma transport coefficients for $\Gamma\lesssim 10$ employing the fast and accurate method of effective potentials developed by Baalrud and Daligault \cite{Daligault1, Daligault}.
   Therefore, the presented work provides an important theoretical basis for the further investigation of the aforementioned physical properties of dense quantum plasmas with strongly coupled ions.
    
 \section*{Acknowledgments}
    
Zh. Moldabekov thanks the funding from
the German Academic Exchange Service (DAAD).
   This work has been supported by the Deutsche Forschungsgemeinschaft via 
    project BO1366/10, and  the Ministry 
of Education and Science of the Republic of Kazakhstan via the grant  BR05236730 ``Investigation of fundamental problems of physics of plasmas and plasma-like media'' (2018).

\section*{Appendix A: Numerical solution of the Ornstein-Zernike equation}

   \begin{figure}
  \includegraphics[width=0.4\textwidth]{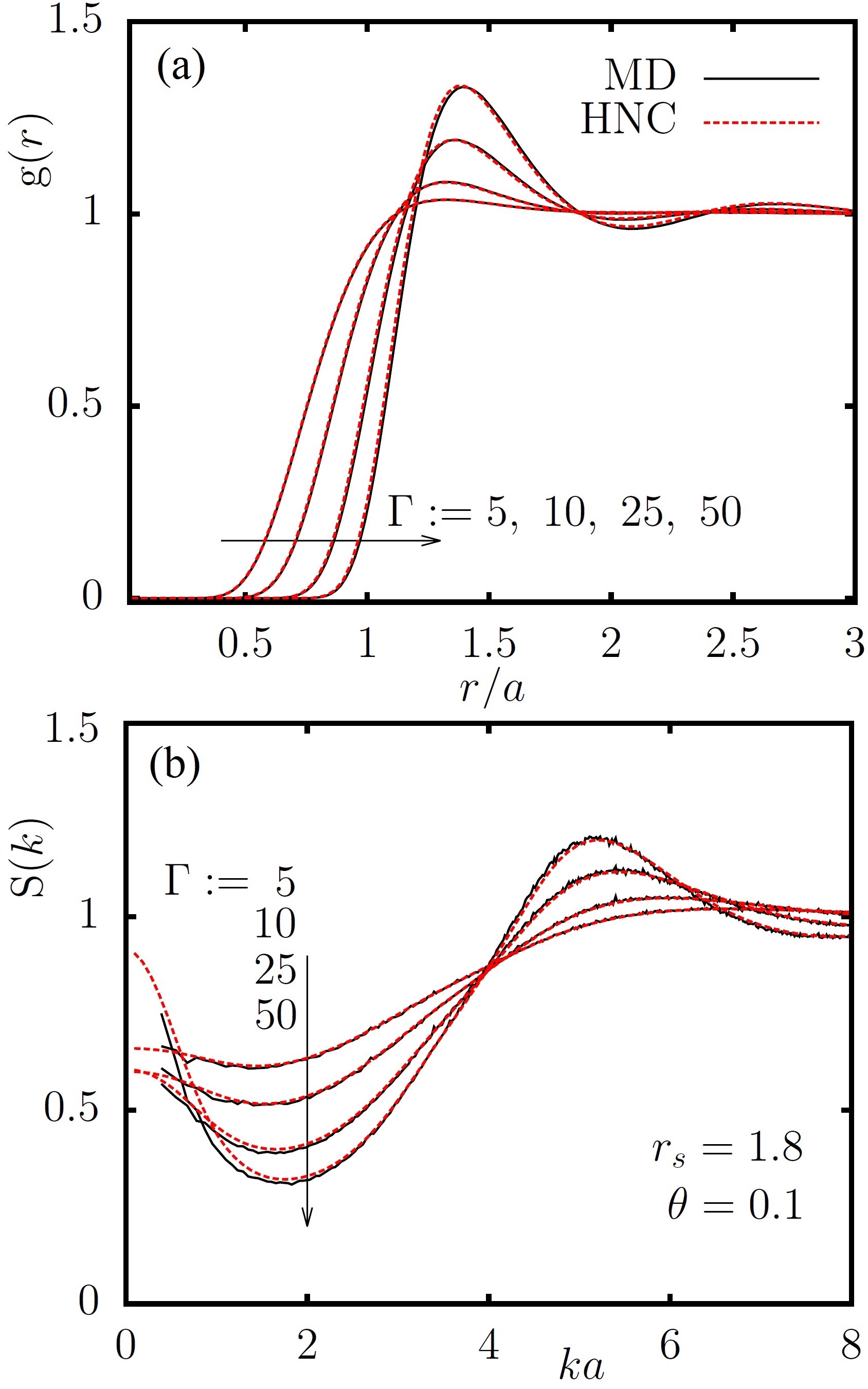}
 \caption{Radial pair distribution function and structure factor of ions interacting through the STLS screened potential computed by the HNC and MD simulations at $r_s=1.8$ and $\theta=0.1$ 
 for different coupling parameters.}
 \label{fig:MD2}
 \end{figure}
 
  The HNC equation  was solved by employing the method of Springer, Pokrant, and Stevens \cite{Pokrant}. The idea of the latter is to rewrite the Ornstein-Zernike equation with the closure relation (\ref{eq:closure}) as
\begin{align}
 \tilde N_s&=\tilde c/(1-n_0\tilde c)-\tilde c_s,\\
 g(r)&=\exp [N_s(r)-u_s(r)],\\
 c_s(r)&=g(r)-1-N_s(r),
\end{align}
where $N(r)=h(r)-c(r)$, and
\begin{align}
 u_s(r)&=u(r)-u_l(r),\\
 c_s(r)&=C(r)+u_l(r),\\
 N_s(r)&=N(r)-u_l(r).
\end{align}
Here $u_l$ is an, in principle,  arbitrary  function that can be used to accelerate  convergence.

For the one-component plasma (OCP), i.e., $\kappa=0$, Ng found that in the liquid state a quick convergence is facilitated by the choice \cite{Ng}
\begin{equation}
 u_l(r)=\frac{\Gamma}{r}\,{\rm erf}(\alpha r),
\end{equation}
where $\alpha=1.08$,  $\kappa=k_s a$ is the screening parameter, and the distance is given in units of $a$.
 
 In the case of the Yukawa one-component plasma (YOCP) with $1<\kappa<2$ and $\Gamma<100$,  the function $u_l$ is not needed for the convergence of the iterations, i.e., $u_l=0$.
However, at $\kappa<1$, the proper choice of $u_l$ is found to be 
\begin{equation}\label{eq:ul}
 u_l(r)=\frac{\Gamma}{r}\left[\exp (-\kappa r)-\exp (-\alpha r)\right],
\end{equation}
where $\alpha=2.16$. Moreover, making use of $u_l$ in the form given by Eq.~(\ref{eq:ul}) improves the convergence of the HNC calculations when the STLS-screened, the RPA-screened and the QMC-based potentials are implemented.

\section*{Appendix B: Verification of the HNC results by MD simulations} 

 As it was demonstrated in the main text, the HNC calculations provide an accurate description of the ionic structural properties, up to $\Gamma_{\rm eff}=10$ and, even at $\Gamma_{\rm eff}=100$,
 correctly capture relative differences when different potentials are used (see Fig. \ref{fig:Geff_HNC}).
 However, for completeness of the study we have checked the findings from the HNC calculations by performing MD simulations based on the Langevin equation of motion.
  The number of ions in the MD simulation is set equal to $N=1000$. The radial pair distribution function and the static structure factor have been calculated separately. 
 
 
 The MD simulations have confirmed our findings 
 from the solution of the  Ornstein-Zernike equation with the HNC closure. This is illustrated in Fig.~\ref{fig:MD2}, where $g(r)$ and $S(k)$ are calculated at $\theta=0.1$ and $r_s=1.8$, for $\Gamma=5,~10,~25$ and 50. 
 From this figure, very good agreement between the MD and the HNC calculations using the STLS-screened potential can be seen. 
 We stress that, at the considered parameters, the HNC works very well, even at $\Gamma=50$. The reason for this is the strong screening, as it was discussed in Sec.~\ref{s:HNC}, for the Yukawa potential.

\end{document}